\documentclass[journal,10pt,a4paper,twoside]{IEEEtran}

\ifCLASSOPTIONdraftcls
	\newcommand{\figheight}{85mm} 
	
\else
  \newcommand{\figheight}{70mm}
  
\fi

\usepackage{ifthen}

\usepackage{amsmath}
\usepackage{amssymb}
\usepackage{amsfonts}
\usepackage{graphicx}
\usepackage{psfrag}
\usepackage{paralist}
\usepackage{times}
\usepackage{amsbsy}
\usepackage{latexsym,bm}
\usepackage{ifsym}
\usepackage{epsfig}
\usepackage{color}
\usepackage{paralist}
\usepackage{psfrag}
\usepackage{stfloats}

\usepackage{amsthm}
\newtheorem{thm}{Theorem}

\newtheorem{cor}{Corollary}
\newtheorem{lem}{Lemma}
\theoremstyle{definition}

\usepackage{bm}

\newcommand{\ma}[1]{\bm{{#1}}}         

\newcommand{\compl}{\mathbb{C}}        
\newcommand{\real}{\mathbb{R}}         
\newcommand{\integer}{\mathbb{Z}}         

\newcommand{\e}{{\rm e}} 
\renewcommand{\j}{\mathrm{j}} 
\newcommand{\sigman}{\sigma_\mathrm{n}}

\newcommand{\opof}[2]{\mathop{{\rm #1}}\left\{#2\right\}}         

\newcommand{\realof}[1]{\opof{Re}{#1}}           
\newcommand{\imagof}[1]{\opof{Im}{#1}}           
\newcommand{\diagof}[1]{\opof{diag}{#1}}         
\newcommand{\traceof}[1]{\opof{Tr}{#1}}       
\newcommand{\expvof}[1]{\opof{\mathbb{E}}{#1}}   

\newcommand{\vecof}[1]{\opof{vec}{#1}}           

\newcommand{\expof}[1]{{\e}^{#1}}                      

\newcommand{\rd}[1]{#1^{({r})}}
\newcommand{\pd}[1]{#1^{({p})}}
\newcommand{\Rd}[1]{#1^{({R})}}
\newcommand{\RdT}[1]{#1^{({R})^\trans}}
\newcommand{\Rdone}[1]{#1^{({1})}}
\newcommand{\RdoneT}[1]{#1^{({1})^\trans}}

\newcommand{\rdH}[1]{#1^{({r})^\herm}}
\newcommand{\pdH}[1]{#1^{({p})^\herm}}
\newcommand{\rdC}[1]{#1^{({r})^\conj}}

\newcommand{\nc}[1]{#1^{({\rm nc})}}

\newcommand{\ncinv}[1]{#1^{({\rm nc})^{-1}}}

\newcommand{\normof}[2]{\left\|#1\right\|_{#2}}
\newcommand{\twonorm}[1]{\normof{#1}{2}}                  

\newcommand{\inv}{{-1}}          
\newcommand{\conj}{*}          
\newcommand{\trans}{{\rm T}}   
\newcommand{\herm}{{\rm H}}    

\usepackage{color}
\newcommand{\red}[1]{{#1}}
\newcommand{\green}[1]{{#1}}

\title{Deterministic Cram\'{e}r-Rao Bound for Strictly Non-Circular Sources and Analytical Analysis of the Achievable Gains}
\author{Jens Steinwandt$^*$,~\IEEEmembership{Student Member,~IEEE},
		    Florian Roemer,~\IEEEmembership{Member,~IEEE}, \\
		    Martin Haardt,~\IEEEmembership{Senior Member,~IEEE}, and Giovanni Del Galdo,~\IEEEmembership{Member,~IEEE}
\thanks{
Parts of this paper have been published at the {\em IEEE Int. Workshop on Smart Antennas (WSA)}, 
Vienna, Austria, Feb. 2007.
}
\thanks{
This work was partially supported by the International Graduate School on Mobile Communications 
(MOBICOM), Ilmenau, Germany.
}
\thanks{
The authors J.~Steinwandt, F.~Roemer, M.~Haardt, and G.~Del Galdo are with Ilmenau University of Technology,
P.O.~Box 100565, D-98684 Ilmenau, Germany,
e-mail: \{jens.steinwandt, florian.roemer, martin.haardt, giovanni.delgaldo\}@tu-ilmenau.de,
phone: +49 (3677) 69-2613, web: http://www.tu-ilmenau.de/crl and http://www.tu-ilmenau.de/dvt.
}
\thanks{$*$ corresponding author}
}

\begin{document}

\maketitle

\linespread{0.973}

\vspace{-1cm}

%
\begin{abstract}
Recently, several high-resolution parameter estimation algorithms have been developed to 
exploit the structure of strictly second-order (SO) non-circular (NC) signals. They achieve 
a higher estimation accuracy and can resolve up to twice as many signal sources compared 
to the traditional methods for arbitrary signals. In this paper, as a benchmark for these 
NC methods, we derive the closed-form deterministic $R$-D NC Cram\'{e}r-Rao bound (NC CRB) 
for the multi-dimensional parameter estimation \red{of} strictly non-circular (rectilinear) 
signal sources. 
Assuming a separable centro-symmetric $R$-D array, we show that in some special cases,
the deterministic $R$-D NC CRB reduces to the existing deterministic $R$-D CRB for arbitrary 
signals. This suggests that no 
gain from strictly non-circular sources (NC gain) can be achieved in these cases.
For more general scenarios, finding an analytical expression of the NC 
gain for an arbitrary number of sources is very challenging. 
Thus, in this paper, we simplify the derived NC CRB and 
the existing CRB for the special case of two closely-spaced strictly non-circular sources 
captured by a uniform linear array (ULA). Subsequently, we use these simplified CRB 
expressions to analytically compute the maximum achievable asymptotic NC gain for the 
considered two source case. The resulting expression only depends on the various physical 
parameters 
and we find the conditions that provide the largest NC gain for two sources. Our analysis 
is supported by extensive simulation results.
\end{abstract}


\IEEEpeerreviewmaketitle

%
%
\section{Introduction} 
\label{sec:intro}
\IEEEPARstart{T}{he} problem of estimating the parameters of multi-dimensional ($R$-D) signals with 
$R\geq 1$, such as their directions of arrival, directions of departure, frequencies, \red{and} Doppler 
shifts, has been an extensive research area with widely-spread signal processing applications 
in radar, sonar, channel sounding, and wireless communications. Recently, various high-resolution 
parameter estimation algorithms such as NC MUSIC \cite{abeida2006perf}, NC Root-MUSIC 
\cite{charge2001ncroot}, NC Standard ESPRIT \cite{zoubir2003ncesprit}, and NC Unitary ESPRIT 
\cite{haardt2004ncunit,steinwandt2014tsp} have been developed to exploit the structure of signals 
from strictly second-order (SO) non-circular (NC) sources \cite{schreier2010noncirc}. The term 
strictly \red{SO NC} (also called rectilinear) is based on the fact that the non-circularity coefficient of these 
signals is equal to one \cite{schreier2010noncirc}. Examples of digital modulation schemes that 
use such signals \red{include} BPSK, PAM, and ASK. The aforementioned NC algorithms that exploit 
the non-circularity property are known to achieve a higher estimation accuracy and can resolve up 
to twice as many sources \cite{steinwandt2014tsp} compared to the traditional 
methods for arbitrary signals \cite{krim1996decades}. However, \red{the ``NC gain'' achieved by these 
algorithms from estimating strictly non-circular sources} has so far only been quantified through 
simulations. Hence, analytical expressions are highly desirable to study the properties 
of the NC gain under various conditions. As deriving a generic formulation for an arbitrary 
number of sources is very challenging, special cases can be considered to provide insights towards 
devising more general expressions. Based on the first-order performance analysis framework in 
\cite{RHD:14}, the scenario of a single strictly non-circular source for NC Standard ESPRIT was 
analyzed in \cite{steinwandt2014tsp}. It was found that no NC gain can be achieved in this case. A 
first attempt at analytically computing the NC gain of NC Standard ESPRIT for two uncorrelated 
strictly non-circular sources with maximum phase separation was taken in \cite{steinwandt2014twosource}. 
Despite the derivation of a closed-form expression for the NC gain, the considered assumptions in 
\cite{steinwandt2014twosource} are rather restrictive and do not provide the desired comprehensive 
insights.

The performance of high-resolution parameter estimation algorithms is often evaluated by comparing 
them to the deterministic (conditional) and stochastic (unconditional) Cram\'{e}r-Rao bounds (CRBs) 
derived in \cite{stoica1989crb} and \cite{stoica2001crb}, respectively. Whereas the stochastic data 
assumption requires both the signals and the noise to be complex Gaussian-distributed, the deterministic 
model assumes that the signals are arbitrary non-random sequences while only the noise follows a complex 
Gaussian distribution. Both CRBs are equally recognized in the literature. 
For the data model used to describe weak-sense non-circular sources whose non-circularity coefficient 
is between zero and one, a stochastic NC CRB has been derived in \cite{delmas2004nccrb}. 
The follow-up papers \cite{abeida2005nccrb} and \cite{delmas2006nccrb} consider further variations of 
the underlying stochastic model assumption. The stochastic NC CRB in \cite{delmas2004nccrb} was derived 
by extending the original Slepian-Bangs formula for circular complex Gaussian distributions 
\cite{stoica2005spectral} to weak-sense non-circular complex Gaussian distributions. However, this bound 
does not apply \red{in} the case of strict non-circularity, as in the weak-sense case, the real part and the 
imaginary part of the signal can be treated as independent random variables. This is not true 
for strictly non-circular sources, where the real and imaginary parts are linearly dependent. 

In this paper, we derive a closed-form expression of the deterministic $R$-D NC CRB for strictly 
non-circular signals impinging on an arbitrary $R$-D sensor array. The derivation is based on the 
conventional Slepian-Bangs formula, which is still applicable due to the complex Gaussian noise 
assumption. Note that our initial contribution in \cite{roemer2007nccrb} only states the 2-D result 
without providing a proof and further analysis. Based on the devised $R$-D NC CRB and assuming the $R$-D 
array to be separable and centro-symmetric, we show that in the special cases of equal \red{rotation} 
phases and full coherence of \red{all} strictly non-circular signals as well as for a single strictly 
non-circular source, the deterministic $R$-D NC CRB reduces to the existing deterministic $R$-D CRB 
for arbitrary signals \cite{stoica1989crb}. This suggests that no NC gain from strictly non-circular 
sources can be achieved in these \red{special} cases. Note that the single source case of the $R$-D 
NC CRB has been analyzed in \cite{steinwandt2014tsp} for a uniform $R$-D array \red{that contains a 
uniform linear array (ULA) in each mode.} Here, we provide a generalization of this case to 
arbitrarily-formed \red{(non-uniform)} separable and centro-symmetric $R$-D arrays. Furthermore, the 
fact that twice as many sources can be resolved from the strictly non-circular data model is highlighted. 

In our second contribution, we \red{assume 1-D parameter estimation} and simplify the derived deterministic 
NC CRB and the deterministic CRB for the special case of two closely-spaced strictly non-circular sources 
captured by a uniform linear array (ULA). These simplified expressions are subsequently used to analytically 
compute the maximum achievable NC gain, which only depends on the physical parameters, e.g., the number of 
sensors, the SNR, the correlation, the phase separation, \red{and} the location of the phase reference of 
the array. The devised expression \red{is based on a truncated Taylor series expansion for closely-spaced 
sources. }
This is, however, the scenario, where 
high-resolution algorithms are primarily applied. Due to the fact that the NC gain expression 
is very general, the properties of the NC gain are studied in terms of the above-mentioned physical 
parameters. For instance, it is shown that the NC gain is largest if the sources are uncorrelated, 
the phase separation is maximum, and the phase reference is at the array centroid. Under these conditions, 
the two sources entirely decouple and do not influence each other. 

The remainder of this paper is organized as follows: The data model is introduced in Section~\ref{sec:data}. 
In Section~\ref{sec:crbs}, the derivation of the deterministic $R$-D NC CRB is provided while its analysis is 
presented in Section~\ref{sec:analysis}. The asymptotic NC gain for two closely-spaced sources 
is analytically computed in \green{Section~\ref{sec:two}.} Section~\ref{sec:simulations} illustrates and 
discusses the numerical results, and concluding remarks are drawn in \green{Section~\ref{sec:conclusions}.}

\textit{Notation:} We use italic letters for scalars, lower-case bold-face letters for column 
vectors, and upper-case bold-face letters for matrices. The superscripts $^\trans$, $^\conj$, 
$^\herm$, and $^\inv$ denote the transposition, complex conjugation, conjugate transposition, 
and the inversion of a matrix, respectively. The Hadamard product of two matrices $\bm A$ and 
$\bm B$ is represented by $\bm A \odot \bm B$, the Kronecker product is symbolized by $\bm A 
\otimes \bm B$, and the Khatri-Rao product (column-wise Kronecker product) is denoted by $\bm A 
\diamond \bm B$. The operator $\vecof{\bm A}$ stacks the columns of the matrix $\bm A \in\compl^{M 
\times N}$ into a column vector of length $MN \times 1$, the operator $\traceof{\bm A}$ returns 
the trace of the matrix $\bm A$, and $\mathrm{diag}\{\bm a\}$ returns a diagonal matrix with the 
elements of $\bm a$ placed on its diagonal. The matrix $\bm \Pi_M$ is the $M \times M$ exchange 
matrix with ones on its antidiagonal and zeros elsewhere. Also, the vector $\bm 1_M$ denotes the 
$M \times 1$ vector of ones while $\bm 1_{M \times M}$ is the $M \times M$ matrix of ones. Moreover, 
$\realof{\cdot}$ and $\imagof{\cdot}$ extract the respective real and imaginary parts of a complex 
number or a matrix, $|\cdot|$ represents the absolute value of a complex number, and $\expvof{\cdot}$ 
stands for the statistical expectation. 
\section{Data Model}
\label{sec:data}
Let $N$ subsequent time instants of the measurement data sampled on an arbitrary separable $R$-D 
grid\footnote{An $R$-D sampling grid is defined to be separable when it is decomposable into the outer 
product of $R$ one-dimensional sampling grids \cite{RHD:14}.} of size $M_1\times\ldots\times M_R$ 
be represented by a linear superposition of $d$ undamped exponentials in additive noise. The $t$-th 
time snapshot of the observed samples can be modeled as \red{ \cite{haardt1998schur} }

\vspace{-1em}
\small
\begin{align}
x_{m_{1},\ldots,m_{R}}(t) = \sum_{i=1}^d s_i(t) \prod_{r=1}^R \expof{\j k_{m_r}\rd{\mu}_i} + n_{m_{1},\ldots,m_{R}}(t),
\label{rdmodel}
\end{align}
\normalsize
%
%
where $m_r=1,\ldots,M_r$, $t=1,\ldots,N$, $s_i(t)$ denotes the complex amplitude of the $i$-th 
undamped exponential at time instant $t$, and $k_{m_r}$ defines the sampling grid\footnote{ The 
number $k_{m_r}$ represents the coordinates of the sampling grid along the $r$-th mode. In terms 
of the spatial domain, it represents the sensor positions of the array in a $\lambda/2$ sampling 
grid. For a uniform sampling grid, we have $k_{m_r} = m_r - 1$.}. 
Moreover, $\rd{\mu}_i$ is the spatial frequency in the $r$-th mode with $i=1,\ldots,d$ and $r=1, 
\ldots,R$, and $n_{m_{1},\ldots,m_{R}}(t)$ contains the \red{ additive zero-mean circularly 
symmetric complex Gaussian noise} samples with variance $\sigman^2$.

In the array signal processing context, each of the $R$-D exponentials represents a narrow-band 
planar wavefront from stationary far-field sources and the complex amplitudes $s_i(t)$ 
are the source symbols. The objective is to estimate the $Rd$ spatial frequencies $\bm \mu_i = 
[\Rdone{\mu}_i,\ldots,\Rd{\mu}_i]^\trans,~\forall i$, from \eqref{rdmodel}. We will also use the 
notation $\rd{\bm \mu} = [\rd{\mu}_1,\ldots,\rd{\mu}_d]^\trans,~\forall r,$ for the spatial frequencies 
of all sources in the $r$-th mode.

In order to obtain a more compact formulation of \eqref{rdmodel}, we collect the observed samples 
into a measurement matrix $\bm X\in\compl^{M\times N}$ with $M = \prod_{r=1}^R M_r$ by stacking 
the $R$ spatial dimensions along the rows and aligning the $N$ time snapshots as the columns. 
Subsequently, $\bm X$ can be modeled as 
\begin{align}
\bm X = \bm A \bm S + \bm N ~ \in\compl^{M\times N},
\label{model}
\end{align}
where $\bm S \in\compl^{d\times N}$ is the source symbol matrix and $\bm N \in\compl^{M\times N}$ 
contains the noise samples. Furthermore, $\bm A = [\bm a(\bm \mu_1),\ldots,\bm a(\bm \mu_d)] 
\in\mathbb C^{M\times d}$ is referred to as the array steering matrix, which consists of the array 
steering vectors $\bm a(\bm \mu_i)$ defined by
\begin{align}
\bm a(\bm \mu_i)= \bm a^{(1)}\left(\mu_i^{(1)}\right)\otimes\cdots\otimes\bm a^{(R)}\left(\mu_i^{(R)}\right) \in\mathbb C^{M\times 1},
\label{steer}
\end{align}
where $\rd{\bm a}(\rd{\mu}_i)\in\mathbb C^{M_r\times 1}$ is the array steering vector of 
the $i$-th spatial frequency in the $r$-th mode. An alternative expression of $\bm A$ is given by 
\begin{align}
\bm A = \bm A^{(1)} \diamond \bm A^{(2)} \diamond \cdots \diamond \bm A^{(R)},
\label{steermat}
\end{align}
where $\rd{\bm A} = [\rd{\bm a}(\rd{\mu_1}),\ldots,\rd{\bm a}(\rd{\mu_d})] 
\in\mathbb C^{M_r\times d}$ represents the array steering matrix in the $r$-th mode. 

The non-circularity of a random variable can be defined through the non-circularity coefficient 
\cite{schreier2010noncirc}. For every complex random variable $Z$ with zero mean, the non-circularity 
coefficient is given by 
\begin{align}
\kappa = \frac{\expvof{Z^2}}{\expvof{|Z|^2}} = |\kappa| \expof{\j\psi}, \quad 0 \leq |\kappa| \leq 1.
\label{ncrate}
\end{align}
The cases $|\kappa| = 0$ and $0 < |\kappa| < 1$ represent a circularly symmetric random variable and a 
weak-sense non-circular variable, respectively. The case $|\kappa| = 1$ represents a strictly 
non-circular (strict sense or rectilinear) random variable. It can be shown that for $|\kappa| = 1$, 
the real part and the imaginary part of $Z$ are linearly dependent \cite{schreier2010noncirc}, 
i.e., $c_1 \cdot \realof{Z} = c_2 \cdot \imagof{Z}$ for constants $c_1, c_2 \in\real$. 

In the context of \red{array processing}, the assumption of strictly SO non-circular sources requires that 
the complex symbol amplitudes of each source lie on a rotated line in the complex plane. This scenario 
is found, for instance, when real-valued data is transmitted by distinct sources causing different delays 
that result in different phase shifts. In this case, the symbol matrix $\bm S$ can be decomposed as 
\cite{haardt2004ncunit}
\begin{equation}
\bm S = \bm \Psi \bm S_0,
\label{ncsignal}
\end{equation}
where $\bm S_0\in\mathbb R^{d\times N}$ is a real-valued symbol matrix and $\bm \Psi=\mathrm{diag}
\{\expof{\j\varphi_i}\}_{i=1}^d$ contains stationary complex phase shifts on its diagonal that can 
be different for each source.

Using \eqref{ncsignal}, the model \eqref{model} can be written as 
\begin{equation}
\bm X = \bm A \bm \Psi \bm S_0 + \bm N.
\label{model2}
\end{equation}
\section{Deterministic $R$-D NC Cram\'{e}r-Rao Bound}
\label{sec:crbs}
In many applications, estimating the parameters of $R$-D signals with $R \geq 1$, can be of high 
importance. As a benchmark of such estimators, the corresponding CRBs for the multi-dimensional 
parameter estimation case are desirable. In this section, we first review the $R$-D CRB for arbitrary 
multi-dimensional signals and then derive the $R$-D NC CRB for multi-dimensional strictly non-circular 
signals. Additionally, we provide simplified expressions of the respective CRBs for the 1-D parameter 
estimation case.
\subsection{Deterministic $R$-D Cram\'{e}r-Rao Bound}
In the case of arbitrary signals, the set of parameters that needs to be considered for the deterministic 
$R$-D CRB is given by the angular parameters $\bm \mu = [\RdoneT{\bm \mu}, \ldots, \RdT{\bm \mu}]^\trans 
\in\real^{Rd \times 1}$, the real part and the imaginary part of the symbols $\bm s = \vecof{\bm S} \in\compl^{Nd 
\times 1}$, and the noise power $\sigman^2$. For this parameter set \red{that contains} a total of $(2N + R)d + 1$ 
parameters, the deterministic CRB matrix in the $R$-D parameter estimation case was derived in \cite{stoica1989crb}. 
Its closed-form expression is given by 
\begin{align}
\bm C = \frac{\sigman^2}{2N} \cdot \realof{\left(\bm D^\herm \bm \Pi^\perp_{\bm A} \bm D \right) \odot 
\RdT{\hat{\bm R}}_S }^\inv \in\real^{Rd \times Rd},
\label{rdcrb}
\end{align}
where 
\begin{align}
\bm \Pi^\perp_{\bm A} = \bm I_M - \bm A \left(\bm A^\herm \bm A \right)^\inv \bm A^\herm \in\compl^{M \times M} 
\end{align}
and 
\begin{align}
\bm D = \begin{bmatrix} \Rdone{\bm D} & \cdots & \Rd{\bm D} \end{bmatrix} \in\compl^{M\times Rd} 
\label{Drd}
\end{align}
with $\rd{\bm D} = [\rd{\bm d}_1,\ldots,\rd{\bm d}_d] \in\compl^{M\times d},~r=1,\ldots,R$, contains the partial 
derivatives of $\bm A$ with respect to \red{the components of} $\bm \mu_i,~i=1,\ldots,d$, in the $r$-th mode. The 
vectors $\rd{\bm d}_i$ are given by $\rd{\bm d}_i = \partial \bm a(\bm \mu_i)/\partial \rd{\mu}_i,~\forall i$. 
Writing $\bm a_i$ instead of $\bm a(\bm \mu_i)$ to simplify the notation and using \eqref{steer}, we obtain
\begin{align}
\rd{\bm d}_i = \bm a_i^{(1)}\otimes\cdots\otimes \bm a_i^{(r-1)} \otimes \rd{\tilde{\bm d}}_i \otimes \bm a_i^{(r+1)} 
\otimes \cdots \otimes \bm a_i^{(R)},
\label{derivrd}
\end{align}
where $\rd{\tilde{\bm d}}_i = \partial \rd{\bm a}_i/\partial \rd{\mu_i}$. 
Moreover, $\Rd{\hat{\bm R}}_S = \bm 1_{R \times R} \otimes \hat{\bm R}_S$ contains the estimated signal covariance matrix 
$\hat{\bm R}_S = \bm \Psi^\conj \hat{\bm R}_{S_0} \bm \Psi$, where the real-valued sample covariance matrix 
$\hat{ \bm R}_{S_0}$ is given by $\hat{\bm R}_{S_0} = \bm S^{}_0\bm S_0^\trans /N$. Note that $\hat{\bm R}_{S_0}$ 
can be written in matrix form as
\begin{align}
	\hat{\bm R}_{S_0} & =  
	     \begin{bmatrix}
	         \hat{P}_1 & \hat{\rho}_{1,2} \sqrt{ \hat{P}_1 \hat{P}_2} & \ldots & \hat{\rho}_{1,d} \sqrt{ \hat{P}_1 \hat{P}_d} \\
	         \hat{\rho}_{2,1} \sqrt{ \hat{P}_1 \hat{P}_2} & \hat{P}_2 & \ldots & \hat{\rho}_{2,d} \sqrt{ \hat{P}_2 \hat{P}_d} \\
	         \vdots & \vdots & \ddots & \vdots \\
	         \hat{\rho}_{d,1} \sqrt{ \hat{P}_1 \hat{P}_d} & \hat{\rho}_{d,2} \sqrt{ \hat{P}_2 \hat{P}_d} & \ldots & \hat{P}_d 
	     \end{bmatrix},
	     \notag
\end{align}
where $\hat{P}_i = \twonorm{\bm s_{0_i}}^2 /N$ is the empirical source power of the $i$-th source and 
$\bm s_{0_i}^\trans \in\real^{1 \times N}$ is the $i$-th row of $\bm S_0$. Furthermore, the \green{empirical} correlation 
coefficients $\hat{\rho}_{i,j}$ that represent the \red{empirical} correlation between the $i$-th and 
the $j$-th source \red{vector} are defined by  
\begin{align}
\hat{\rho}_{i,j} = \frac{1}{N} \cdot \frac{\bm s_{0_i}^\trans \bm s^{}_{0_j}}{\sqrt{\hat{P}_1 \hat{P}_2}},~~ \forall \; i \neq j,~~ i,j = 1,\ldots,d.
\label{corrdef}
\end{align}
Note that $\bm{\hat{R}}_{S_0}$ is symmetric such that $\hat{\rho}_{i,j} = \hat{\rho}_{j,i}$.

In the special case of 1-D parameter estimation, the array steering matrix $\bm A$ reduces to $\bm A = 
[\bm a(\mu_1),\ldots,\bm a(\mu_d)]\in\compl^{M \times d}$ and the deterministic CRB matrix in 
\eqref{rdcrb} simplifies to
\begin{align}
\bm C = \frac{\sigman^2}{2N} \cdot \realof{\left(\bm D^\herm \bm \Pi^\perp_{\bm A} \bm D \right) 
\odot \hat{\bm R}_S^\trans }^\inv \in\real^{d \times d},
\label{crb}
\end{align}
where $\bm D$ becomes
\begin{align}
\bm D = \begin{bmatrix} \bm d_1 & \cdots & \bm d_d \end{bmatrix} \in\compl^{M \times d}
\label{Dmat}
\end{align}
with $\bm d_i = \partial \bm a(\mu_i)/\partial \mu_i,~\forall \;i$.
\subsection{Deterministic $R$-D NC Cram\'{e}r-Rao Bound}
In contrast to the case of arbitrary signals, the set of parameters for the strictly non-circular source 
model in \eqref{model2} is given by the angular parameters $\bm \mu \in\real^{Rd \times 1}$, the real-valued 
symbols $\bm s_0 = \vecof{\bm S_0} \in\real^{Nd \times 1}$, the rotation phase angles $\bm \varphi \in\real^{d 
\times 1}$, and the noise power $\sigman^2$. Thus, the number of parameters is now equal to $(N+R+1)d + 1$, which 
requires the derivation of a new CRB for this parameter set.

The resulting closed-form expression for the deterministic NC CRB matrix $\nc{\bm C}$ in the 
$R$-D case is stated in the following theorem:
\begin{thm}
The $R$-D deterministic NC CRB matrix $\nc{\bm C}$ for strictly non-circular sources is given by
\begin{align}
&\nc{\bm C} = \frac{\sigman^2}{2N} \cdot \Big\{\left(\bm{G}_2 - \bm{G}_1 \bm{G}_0^\inv \bm{G}_1^\trans \right) \odot \Rd{\hat{\bm R}}_{S_0} \notag \\ 
&+ \left[ \left( \bm{G}_1 \bm{G}_0^\inv \bm{H}_0 \right) \odot \Rd{\hat{\bm R}}_{S_0} \right] \!\left[ \left( \bm{G}_0 - \bm{H}_0^\trans \bm{G}_0^\inv \bm{H}_0 
\right) \odot \Rd{\hat{\bm R}}_{S_0} \right]^\inv \notag\\
& \cdot \left[ \left(\bm{H}_1^\trans -  \bm{H}_0^\trans \bm{G}_0^\inv \bm{G}_1^\trans \right) \odot \Rd{\hat{\bm R}}_{S_0} \right] 
+ \left[ \bm{H}_1 \odot \Rd{\hat{\bm R}}_{S_0} \right] \notag\\
& \cdot \left[ \bm{G}_0 \odot \Rd{\hat{\bm R}}_{S_0} \right]^\inv \!\!\cdot \left[ \left( \bm{H}_0^\trans \bm{G}_0^\inv \bm{G}_1^\trans \right) \odot 
\Rd{\hat{\bm R}}_{S_0} \right] + \left[ \bm{H}_1 \odot \Rd{\hat{\bm R}}_{S_0} \right] \notag\\ 
& \cdot \left[ \bm{G}_0 \odot \Rd{\hat{\bm R}}_{S_0} \right]^\inv
\cdot \left[ \left( \bm{H}_0^\trans \bm{G}_0^\inv \bm{H}_0 \right) \odot \Rd{\hat{\bm R}}_{S_0} \right] \notag\\
& \cdot \left[ \left( \bm{G}_0 - \bm{H}_0^\trans \bm{G}_0^\inv \bm{H}_0 \right) \odot \Rd{\hat{\bm R}}_{S_0} \right]^\inv \!\!\!\cdot \!\left[ \left( \bm{H}_0^\trans \bm{G}_0^\inv \bm{G}_1^\trans \right) \odot \Rd{\hat{\bm R}}_{S_0} \right] \notag\\
&- \left[ \bm{H}_1 \odot \Rd{\hat{\bm R}}_{S_0} \right] \cdot \left[ \left( \bm{G}_0 - \bm{H}_0^\trans \bm{G}_0^\inv \bm{H}_0 
\right) \odot \Rd{\hat{\bm R}}_{S_0} \right]^\inv \notag\\ 
&\cdot \left[ \bm{H}_1^\trans \odot \Rd{\hat{\bm R}}_{S_0} \right] \Big\}^\inv \in\real^{Rd \times Rd},
\label{eqn_thekillercrb}
\end{align}
where $\Rd{\hat{\bm R}}_{S_0} = \bm 1_{R \times R} \otimes \hat{\bm R}_{S_0}$ and the matrices $\bm G_n$ and 
$\bm H_n,~n=0,1,2$, are defined as
\begin{align}
\bm G_0 & = \realof{\bm \Psi^\conj \bm A^\herm \bm A \bm \Psi} \in\real^{d\times d}, \label{g0rd}\\
\bm H_0 & = \imagof{\bm \Psi^\conj \bm A^\herm \bm A \bm \Psi} \in\real^{d\times d}, \label{h0rd}\\
\bm G_1 & = \realof{(\bm I_R \otimes \bm \Psi^\conj) \bm D^\herm \bm A \bm \Psi} \in\real^{Rd \times d}, \label{g1rd}\\
\bm H_1 & = \imagof{(\bm I_R \otimes \bm \Psi^\conj) \bm D^\herm \bm A \bm \Psi} \in\real^{Rd \times d}, \label{h1rd}\\
\bm G_2 & = \realof{(\bm I_R \otimes \bm \Psi^\conj) \bm D^\herm \bm D (\bm I_R \otimes \bm \Psi)} \in\real^{Rd \times Rd}
\label{g2rd}
\end{align}
and $\bm A$ and $\bm D$ are given by \eqref{steermat} and \eqref{Drd}, respectively.
\label{thm:nccrb}
\end{thm}
\begin{IEEEproof}
The proof is given in Appendix \ref{app:nccrb}.
\end{IEEEproof}
It should be highlighted that the assumption of the $R$-D array to be separable is not required for the 
derivation of \eqref{eqn_thekillercrb} in Appendix \ref{app:nccrb}. In fact, \eqref{eqn_thekillercrb} 
is valid for arbitrarily formed $R$-D \red{arrays\footnote{These also include non-separable arrays such as 
cross-arrays and L-shaped arrays.}}, where the columns of $\bm A$ and $\bm D$ are represented 
accordingly. However, the separability assumption simplifies the further analysis and helps with the 
presentation of our results in the following sections.

In analogy to the 1-D parameter estimation case of the CRB for arbitrary signals in \eqref{crb}, the deterministic 
1-D NC CRB matrix is stated in the corollary: 
\begin{cor}
The deterministic 1-D NC CRB is given by \eqref{eqn_thekillercrb}, where $\Rd{\hat{\bm R}}_{S_0}$ reduces 
to $\hat{\bm R}_{S_0}$ and $\bm G_1$, $\bm H_1$, and $\bm G_2$ simplify to
\begin{align}
\bm G_1 & = \realof{\bm \Psi^\conj \bm D^\herm \bm A \bm \Psi} \in\real^{d \times d}, \\
\bm H_1 & = \imagof{\bm \Psi^\conj \bm D^\herm \bm A \bm \Psi} \in\real^{d \times d}, \\
\bm G_2 & = \realof{\bm \Psi^\conj \bm D^\herm \bm D \bm \Psi} \in\real^{d \times d} 
\end{align}
with $\bm A$ and $\bm D$ being defined in \eqref{crb} and in \eqref{Dmat}.
\label{cor:rdnccrb}
\end{cor}
\section{Analysis of the Deterministic $R$-D NC CRB}
\label{sec:analysis}
In this section, we discuss interesting special cases and properties of the derived $R$-D NC CRB, 
where the $R$-D array is assumed to be separable and centro-symmetric for simplicity. Specifically, 
we investigate the two cases of equal rotation phases and full coherence for an arbitrary number of 
strictly non-circular signals before we focus on the single source case ($d=1$). \red{It is shown that 
in these special cases, the deterministic $R$-D NC CRB reduces to the $R$-D CRB. Furthermore, 
we also analyze the maximum number of resolvable NC sources. }

For our analysis, we first refine the model in \eqref{model2}. Assuming the $R$-D array to be 
centro-symmetric, i.e., it is symmetric with respect to its centroid, its array steering matrices 
$\rd{\bm A}$ from \eqref{steermat} satisfy \cite{haardt1995unitesprit}
\begin{align}
\bm \Pi_{M_r} \rdC{\bm A} = \rd{\bm A} \rd{\bm \Delta}_\mathrm{c} ~~\forall \;r,
\label{centro}
\end{align} 
where $\rd{\bm \Delta}_\mathrm{c} \in\compl^{d \times d}$ is a unitary diagonal matrix that depends 
on the phase reference. If the $r$-mode array centroid is chosen as the phase reference 
\cite{haardt1995unitesprit}, we have $\rd{\bm \Delta}_\mathrm{c} = \bm I_d$. The phase reference 
$\rd{\delta}$ along the $r$-th mode can be defined by
\begin{align}
\rd{\delta} = \frac{1}{M_r} \sum_{m_r=1}^{M_r} k_{m_r}.
\label{phasedef}
\end{align}
Note that $\rd{\delta}$ is a property of the array and independent of $\rd{\mu}_i$. Using \eqref{centro} 
and \eqref{phasedef}, we can decompose the $r$-mode array steering matrix $\rd{\bm A}$ with an arbitrary 
phase reference along the $r$-th mode as 
\begin{align}
\rd{\bm A} = \rd{\bar{\bm A}} \rd{\bm \Delta}.
\label{arrayphaserd}
\end{align}
The matrix $\bar{\bm A} \in\compl^{M_r \times d}$ is the array steering matrix whose phase reference is 
located at the centroid of the $r$-th mode such that for $\rd{\bar{\bm A}}$, from \eqref{centro}, the 
identity $\rd{\bar{\bm A}} = \bm \Pi_{M_r} \rdC{\bar{\bm A}}$ holds. Furthermore, the diagonal matrix 
$\rd{\bm \Delta} = \mathrm{diag}\big\{\e^{\j \rd{\delta} \rd{\mu}_i} \big\}_{i=1}^d$ contains the shifts 
of the phase reference for each $\rd{\mu}_i$. By inserting \eqref{arrayphaserd} into \eqref{centro}, we 
can easily establish the relation $\rd{\bm \Delta}_\mathrm{c} = \rdC{\bm \Delta} \rdC{\bm \Delta}$. Thus, 
if the actual phase reference is at the array centroid of the $r$-th mode, we have $\rd{\delta} = 0$, 
$\rd{\bm \Delta} = \rd{\bm \Delta}_\mathrm{c} = \bm I_d$, and consequently $\rd{\bm A} = \rd{\bar{\bm A}}$. 

Based on \eqref{arrayphaserd}, we can rewrite $\bm A$ in \eqref{steermat} as 
\begin{align}
\bm A & = \left( \bar{\bm A}^{(1)} \bm \Delta^{(1)} \right) \diamond \left( \bar{\bm A}^{(2)} \bm \Delta^{(2)} \right) \diamond \cdots 
\diamond \left( \bar{\bm A}^{(R)} \bm \Delta^{(R)} \right) \notag\\
&= \bar{\bm A} \bm \Delta,
\label{steermatphase}
\end{align}
where $\bar{\bm A} = \bar{\bm A}^{(1)} \diamond \bar{\bm A}^{(2)} \diamond \cdots \diamond \bar{\bm A}^{(R)} 
\in\compl^{M \times d}$ and $\bm \Delta = \bm \Delta^{(1)} \cdot \bm \Delta^{(2)} \cdot \ldots \cdot
\bm \Delta^{(R)}\in\compl^{d \times d}$.

Inserting \eqref{steermatphase} into the expression for strictly non-circular signals in \eqref{model2}, 
we obtain
\begin{align}
\bm X = \bar{\bm A} \bm \Delta \bm \Psi \bm S_0 + \bm N = \bar{\bm A} \bm \Phi \bm S_0 + \bm N,
\label{model3}
\end{align}
where we have defined $\bm \Phi = \bm \Delta \bm \Psi = \diagof{\e^{\j (\varphi_i + \delta_i )}}_{i=1}^d$ 
with $\delta_i = \sum_{r=1}^R \rd{\delta} \rd{\mu}_i$. 
\subsection{Sources with Equal Phases}
\label{sec:equalphase}
An interesting special case of the model \eqref{steermatphase} occurs when the phase references in each 
of the $R$ modes coincide with the centroid of the $R$-D array, i.e., $\rd{\delta} = 0~\forall \;r$ such 
that $\bm \Delta = \bm I_d$ and $\bm A = \bar{\bm A}$, and, at the same time, the rotation phase angles 
for all $d$ sources are the same\footnote{The same behavior applies to the more \red{general} case of equality 
modulo $\pi$, i.e., $\varphi_i = \varphi + k_i \cdot \pi$, $k_i \in \integer$ for $i=1,2,\ldots,d$. For 
simplicity of presentation, we assume the angles to be equal, \green{this} generalization is however 
straightforward.}, i.e., $\varphi_i = \varphi~\forall \;i$. Hence, we have
\begin{align}
	\bm \Phi = \bm \Psi = \e^{\j \varphi} \bm I_d.
\end{align}
Under these assumptions, the matrices $\bm G_n$, $n=0,1,2$, can be expressed as
\begin{align*}
	\bm G_0 & = \realof{\e^{-\j \varphi} \bm I_d \bm A^\herm \ma{A} \ma{I}_d \e^{\j \varphi}}
	= \realof{\ma{A}^\herm \ma{A}} = \ma{A}^\herm \ma{A} \\
	\ma{G}_1 & = \realof{\e^{-\j \varphi} \ma{I}_{Rd} \bm D^\herm \ma{A} \ma{I}_d \e^{\j \varphi}} 
	= \realof{\bm D^\herm \ma{A}} = \bm D^\herm \bm A \\
	\ma{G}_2 & = \realof{\e^{-\j \varphi} \ma{I}_{Rd} \bm D^\herm \bm D \ma{I}_{Rd} \e^{\j \varphi}}
	= \realof{\bm D^\herm \bm D} = \bm D ^\herm \bm D
\end{align*}
while the matrices $\ma{H}_n$ evaluate to zero. The proof that the matrices $\bm A^\herm \bm A \in\real^{d \times d}$, 
$\bm D^\herm \bm A \in\real^{Rd \times d}$, and $\bm D^\herm \bm D \in\real^{Rd \times Rd}$ are 
real-valued can be found in Appendix \ref{app:real}.

Using these observations, all terms in (\ref{eqn_thekillercrb}) containing $\ma{H}_0$ or $\ma{H}_1$
vanish and the $R$-D NC CRB matrix simplifies to 
\begin{align}
\nc{\bm C} & = \frac{\sigman^2}{2N} \cdot \Big\{
   	\left(\bm G_2 - \bm G_1 \bm G_0^\inv \bm G_1^\trans \right) \odot \hat{\bm R}^{(R)}_{S_0} \Big\}^\inv \nonumber \\
         & = \frac{\sigman^2}{2N} \cdot \bigg\{\big(\bm D^\herm \bm \Pi_{\bm A}^\perp \bm D \big) 
         \odot \hat{\bm R}_{S}^{(R)^\trans} \bigg\}^\inv \notag \\
   			 & = \bm C, 
\label{eqn_psiI_ceq}
\end{align}
where we have used the fact that $\hat{\bm R}^{(R)}_S = \hat{\bm R}^{(R)}_{S_0} = \hat{\bm R}_{S_0}^{(R)^\trans}$ 
for $\bm \Psi = \e^{\j \varphi} \bm I_d$. From \eqref{eqn_psiI_ceq}, it is evident that the $R$-D NC CRB reduces 
to the $R$-D CRB if the phase reference is at the $R$-D array centroid and the rotation phase angles of the sources 
are equal. This suggests that no gain from strictly non-circular sources can be achieved in this case.  
\subsection{Coherent Sources}
\label{sec:coherent}
In this section, the case of full coherence is discussed, i.e., the correlation coefficients $\hat{\rho}_{i,j}$ 
between all pairs of sources are given by \red{$|\hat{\rho}_{i,j}| = 1 ~\forall \;i,j$.} For simplicity, we 
assume \red{that} all the sources have unit power, i.e., $\hat{P}_i = 1 ~\forall \;i$. Under these assumptions, the sample 
covariance matrix takes the form $\hat{\bm R}_{S_0} = \bm 1_{d \times d}$ such that $\Rd{\hat{\bm R}}_{S_0} = 
\bm 1_{Rd \times Rd}$. Hence, all the Hadamard products with $\Rd{\hat{\bm R}}_{S_0}$ in the $R$-D NC CRB matrix 
in \eqref{eqn_thekillercrb} \red{can be omitted} and the remaining parts \red{are} arranged in the following form
\begin{align}
&\!\!\!\frac{\sigman^2}{2N} \cdot \ncinv{\bm C} \!\! =  
\ma{G}_2 - \ma{G}_1 \!\left[ \ma{G}_0^\inv + \ma{G}_0^\inv \ma{H}_0 \tilde{\bm G}^\inv
  \ma{H}_0^\trans \ma{G}_0^\inv \right] \ma{G}_1^\trans \nonumber \\
&~~~ + \ma{H}_1 \ma{G}_0^\inv \ma{H}_0^\trans \left[ \ma{G}_0^\inv 
     + \ma{G}_0^\inv \ma{H}_0 \tilde{\bm G}^\inv
  \ma{H}_0^\trans \ma{G}_0^\inv \right] \ma{G}_1^\trans \nonumber \\
&~~~ + \ma{G}_1 \ma{G}_0^\inv \ma{H}_0 \tilde{\bm G}^\inv \ma{H}_1^\trans
  - \ma{H}_1 \tilde{\bm G}^\inv \ma{H}_1^\trans \label{ir1} \\
& = \ma{G}_2 - \left(\ma{G}_1 - \ma{H}_1 \ma{G}_0^\inv \ma{H}_0^\trans \right) 
                  \left( \ma{G}_0 - \ma{H}_0 \ma{G}_0^\inv \ma{H}_0^\trans \right)^\inv \ma{G}_1^\trans \nonumber \\
&~~~ - \left(\ma{H}_1 - \ma{G}_1 \ma{G}_0^\inv \ma{H}_0 \right) \!\left( \ma{G}_0 - \ma{H}_0^\trans \ma{G}_0^\inv \ma{H}_0
\right)^\inv \!\!\ma{H}_1^\trans,
\label{eqn_crbcoh}
\end{align}
where in \eqref{ir1}, we have defined $\tilde{\bm G} = \bm G_0 - \bm H_0^\trans \bm G_0^\inv \bm H_0$ and replaced 
the terms in the square brackets by applying the converse of the matrix inversion lemma, yielding the matrix $(\bm G_0 - 
\bm H_0 \bm G_0^\inv \bm H_0^\trans)^\inv$. Note that \eqref{eqn_crbcoh} can be transformed into the block matrix form 
\begin{align} \notag
	\frac{\sigman^2}{2N} \cdot \ncinv{\bm C} & = \ma{G}_2  - \begin{bmatrix} \ma{G}_1 & \ma{H}_1 \end{bmatrix}
	\begin{bmatrix} \ma{G}_0 & \ma{H}_0 \\ \ma{H}_0^\trans & \ma{G}_0 \end{bmatrix}^\inv 
	\begin{bmatrix} \ma{G}_1^\trans \\ \ma{H}_1^\trans \end{bmatrix},
\end{align}
which represents a very interesting simplification of the original expression.

In the next step, we rewrite the $R$-D CRB matrix for arbitrary signals in \eqref{rdcrb} for the case of 
coherent sources in a similar form. Under the aforementioned assumptions, the $R$-D sample covariance 
matrix is given by $\Rd{\hat{\bm R}}_S = \bm 1_{R \times R} \otimes (\bm \Psi^\conj \bm 1_{d \times d} 
\bm \Psi) = (\bm I_R \otimes \bm \Psi^\conj)\bm 1_{Rd \times Rd}(\bm I_R \otimes \bm \Psi)$. Hence, we 
simplify the original form of the CRB in \eqref{crb} into
\begin{align}
  &\frac{\sigman^2}{2N} \cdot \bm C^\inv \!= \mathrm{Re}\Big\{ \!\big(\bm D^\herm \bm \Pi^\perp_{\bm A} \bm D \big) \Big. \notag\\
  & \Big. \quad\quad\quad\quad\quad\quad\quad \odot \big( (\bm I_R \otimes \bm \Psi^\conj)\bm 1_{Rd \times Rd}(\bm I_R \otimes \bm \Psi) \big) \! \Big\}  \\
 	& = \mathrm{Re}\Big\{ (\bm I_R \otimes \bm \Psi^\conj) \bm D^\herm \bm D (\bm I_R \otimes \bm \Psi) - 
 	(\bm I_R \otimes \bm \Psi^\conj) \bm D^\herm \bm A \bm \Psi \Big. \notag \\
 	& \Big. \quad\quad\quad\quad\quad~ \cdot \left(\bm \Psi^\conj \bm A^\herm \bm A 
 	\bm \Psi\right)^\inv  \bm \Psi^\conj \bm A^\herm \bm D (\bm I_R \otimes \bm \Psi) \Big\} \label{addpsi} \\
  & = \mathrm{Re}\Big\{ \ma{G}_2 + \j \ma{H}_2 - \left( \ma{G}_1 + \j \ma{H}_1 \right)  \Big. \notag\\ 
  & \Big. \quad\quad\quad\quad\quad~ \cdot \left( \ma{G}_0 + \j \ma{H}_0 \right)^\inv ( \ma{G}_1^\trans - \j \ma{H}_1^\trans ) \Big\}, 
  \label{crb_mod}
\end{align}
where we have introduced additional matrices $\bm \Psi$ in \eqref{addpsi} by noting that $\bm \Psi \bm 
\Psi^\conj = \bm I_d$. To proceed we require the following lemma:
\begin{lem}
   The inverse of a full rank complex-valued matrix $\bm C = \bm A + \j \bm B \in \compl^{n \times n}$ 
   with the real part $\bm A \in \real^{n \times n}$ and the imaginary part $\bm B \in \real^{n \times n}$
	 can be split into its real part and its imaginary part as follows:
   \begin{align} \notag
 	    \bm C^\inv 
 	    & = \left( \ma{A} + \ma{B} \ma{A}^\inv \ma{B} \right)^\inv
 	        - \j \ma{A}^\inv \ma{B} \left( \ma{A} + \ma{B} \ma{A}^\inv \ma{B} \right)^\inv \notag
   \end{align}
   if $\bm A$ and $\left( \bm A + \bm B \bm A^\inv \bm B \right)$ are invertible.
   \label{lem_invcompl}
\end{lem}
\begin{IEEEproof}
To prove this lemma, it is sufficient to multiply $\bm C$ with $\bm C^\inv$ and show that the result 
is the identity matrix.
\end{IEEEproof}
Applying Lemma \ref{lem_invcompl} to \eqref{crb_mod}, we split $\bm C^\inv$ into its real and imaginary 
part. After some elementary operations and using the fact that $\bm H_0^\trans = -\bm H_0$, we obtain 
equation \eqref{eqn_crbcoh} and consequently, we have $\nc{\bm C} = \bm C$. Thus, both $R$-D CRBs become 
equal if all the sources are coherent. Note that this result is valid for arbitrary $R$-D arrays as the 
assumptions of separability and centro-symmetry were not used in the derivation. Analogously to the 
special case considered in the previous subsection, our findings suggests that no NC gain can be achieved 
for coherent sources.
\subsection{Single Source Case}
\label{sec:single}
The expression for the deterministic $R$-D NC CRB is formulated in terms of the matrices $\bm A$, 
$\bm \Psi$, $\bm D$, and the sample covariance matrix $\hat{\bm R}_{S_0}$. Consequently, it 
provides no explicit insights into the parameters of physical significance, e.g., the number of sensors 
$M$, the correlation coefficient $\rho$, the source separation. Knowing how the CRB scales with 
these parameters can facilitate array design decisions on the number of required sensors to achieve 
a certain performance under specific conditions. As establishing a generic formulation for an arbitrary 
number of sources is very challenging, we can, however, consider special cases such as the single 
source case. Note that this scenario of the $R$-D NC CRB has been analyzed in \cite{steinwandt2014tsp} 
for a uniform $R$-D array containing a uniform linear array (ULA) in the $r$-th mode and we found that 
no NC gain can be achieved in this case. Here, we provide a generalization of the previous results and 
simplify the $R$-D NC CRB for \green{non-uniform centro-symmetric and separable} $R$-D arrays. 

So far, we have shown that, from the $R$-D NC CRB, no NC gain can be obtained if the sources have the 
same rotation phase while the phase reference is at the array centroid or if the sources are coherent. 
As the single source case is a special case of each of these two properties, i.e., $\bm \Psi = 
\e^{\j\varphi}$ or $\Rd{\hat{\bm R}}_{S_0} = \bm 1_{R \times R}$, we can directly conclude 
that the $R$-D NC CRB and the $R$-D CRB must be equal for this case as well, which is in line with our 
results in \cite{steinwandt2014tsp}.

The simplified expression of the deterministic $R$-D NC CRB for a single strictly non-circular source 
is shown in the next theorem:
\begin{thm}
For the case of an $M_1 \times \ldots \times M_R$ ($M$-element) separable $R$-D array with \green{$\rd{\delta} 
= 0~\forall~r$, i.e., the phase reference of the centro-symmetric array is at the centroid}, and a single 
strictly non-circular source ($d=1$), the deterministic $R$-D NC CRB can be simplified to
\begin{align}
\nc{\bm C} = \diagof{\left[ {\nc{C}}^{(1)},\ldots, {\nc{C}}^{(R)} \right]} \in\real^{R \times R}
\label{nccrb_single}
\end{align}
with
\begin{align}
\rd{{\nc{C}}} = \frac{1}{\hat{\varrho}} \cdot \frac{M_r}{2 M} \cdot \frac{1}{\sum_{m_r=1}^{M_r} k_{m_r}^2}~~ \forall \; r,
\label{nccrb_single1}
\end{align}
where $\hat{\varrho}$ represents the effective SNR $\hat{\varrho} = N \hat{P} / \sigman^2$ 
with $\hat{P}$ being the empirical source power given by $\hat{P} = \twonorm{\bm s_0}^2 / N$ and 
$\bm s_0 \in \real^{N \times 1}$. 
\label{thm:nccrbsingle}
\end{thm}
\begin{IEEEproof}
The proof is given in Appendix \ref{app:nccrbsingle}.
\end{IEEEproof}
For the special case of a uniform $R$-D sampling grid, the $R$-D NC CRB expression from Theorem \ref{thm:nccrbsingle} 
\red{is simplified in the following corollary:}
\begin{cor}
For an $M$-element uniform $R$-D array with an $M_r$-element ULA in the $r$-th mode and a single strictly 
non-circular source ($d=1$), the deterministic NC CRB for the $r$-th mode in \eqref{nccrb_single1} can be 
explicitly expressed as
\begin{align}
\rd{{\nc{C}}} = \frac{1}{\hat{\varrho}}\cdot\frac{6}{M(M_r^2-1)}~~ \forall \; r,
\label{nccrb_single_ula}
\end{align}
where $k_{m_r} = -(M_r -1)/2,\ldots,(M_r - 1)/2$.
\label{thm:nccrbsingleula}
\end{cor}
The expression \eqref{nccrb_single_ula} is in line with our previous developments in \cite{steinwandt2014tsp}.  
Moreover, \eqref{nccrb_single_ula} is equivalent to the result for the single source case of the deterministic 
$R$-D CRB for arbitrary signals derived in \cite{roemer2012perf}. This fact proves our previous claim that no 
improvement in terms of the estimation accuracy can be achieved for a single strictly non-circular source. 
\subsection{Maximum Number of Resolvable Sources}
In the case of arbitrary signals, it is well-known from \cite{stoica1989crb} that the upper limit\footnote{This 
limit is not reached with all array geometries. An example for an array, which can achieve this limit 
is a ULA.} of sources that can be resolved with $M$ sensors is $d=M-1$. However, if the sources are 
strictly non-circular, we can estimate the DOAs of even more sources than sensors available. In this 
section, we establish the conditions under which the deterministic NC CRB is valid for $d \geq M$.

Firstly, it is not difficult to see that the matrices $\bm G_n$ and $\bm H_n$, $n=0, 1, 2$,
can have a rank larger than $M$. For example, the matrix $\bm G_0$ can be rewritten
as
\begin{align}
	\bm G_0 = \realof{\bm \Psi^\conj \bm A^\herm \bm A \bm \Psi}
	= \begin{bmatrix}
	      \bm A \realof{\bm \Psi} \\
	      \bm A \imagof{\bm \Psi} 
	  \end{bmatrix}^\herm  
	  \underbrace{
	  \begin{bmatrix}
	      \bm A \realof{\bm \Psi} \\
	      \bm A \imagof{\bm \Psi} 
	   \end{bmatrix}
	   }_{2M \times d}. 
	   \notag
\end{align}
From this equation, it can be seen that unless the phase matrix $\bm \Psi$ is equal 
to $\bm \Psi = \mathrm{diag}\{\e^{\j\varphi_i}\}_{i=1}^d$ with $\varphi_i = \varphi + k_i \cdot 
\pi,~k_i \in\integer$, i.e., all the rotation phases are equal modulo $\pi$, $\bm G_0$ has 
a rank larger than $M$ if $d > M$. This result complies with the one from Subsection \ref{sec:equalphase}. 
For the matrices $\bm G_1$, $\bm G_2$ as well as $\bm H_n$, $n=0,1$, similar forms are easily 
found. 

Secondly, regarding the additional dependence of the NC CRB on the sample covariance matrix 
$\hat{\bm R}_{S_0}$, we have proven in Subsection \ref{sec:coherent} that the NC CRB reduces to 
the CRB if the sources are coherent. This suggests that for non-coherent sources, the NC CRB is valid for 
$d \geq M$.

Consequently, we can infer for a uniform linear array that if the sources are non-coherent, i.e.,
\begin{equation}
	\red{|\hat{\rho}_{i,j}|} < 1 \quad \forall i \neq j \;\; \mbox{in} \;\; 1, 2, \ldots, d,
\end{equation}
and the rotation phase angles are different, i.e.,
\begin{equation}
	|\varphi_i - \varphi_j| \neq 0 \mod \pi \quad \forall i \neq j \;\; \mbox{in} 
	\;\; 1, 2, \ldots, d, 
\end{equation}
the Fisher information matrix \red{has} full rank and \red{is} invertible as long as $d \leq 2(M-1)$. 
To support our claim, we provide the numerical evaluation shown in Table~\ref{table}, which suggests 
that the condition $d \leq 2(M-1)$ represents an upper limit on the number of sources that is resolvable. 
Therefore, up to twice as many signal sources can be resolved compared to the case of arbitrary signals. 
\section{Achievable NC Gain for Two Sources}
\label{sec:two}
After establishing that according to the $R$-D NC CRB, no NC gain can be attained for a single source, 
the question to be studied is what is the maximum achievable NC gain if at least two sources are not 
fully coherent, their rotation phases are different, and the phase reference is arbitrary. 
Although it is well known that exploiting the properties of strictly non-circular sources can 
provide significant gains in reducing the estimation error, so far, the NC gain could 
only be quantified via simulations. In this section, we analytically compute the maximum achievable NC 
gain associated with strictly non-circular sources. As finding an analytical expression for an arbitrary 
number of sources is an intricate task, we limit our analysis to the case of two closely-spaced strictly 
non-circular sources. The $R$-D CRB for arbitrary source constellations tends to infinity when the source 
separation approaches zero. This is not always true for the $R$-D NC CRB as \red{under} certain conditions, a finite 
value is reached. This observation motivates us to derive simplified expressions of the NC CRB and the 
CRB for the two source case, which are subsequently used to analytically compute the maximum achievable 
NC gain. To obtain generic expressions in terms of the physical parameters, the derivations are based 
on the model in \eqref{model3}.
 
For simplicity, we limit our analysis to the 1-D parameter estimation case and assume a ULA composed of 
$M$ isotropic sensor elements, which is centro-symmetric. The phase reference is located \red{at an arbitrary 
position.} 
For this scenario, the array steering matrix $\bar{\bm A}$ in model \eqref{model3} 
simplifies to 
\begin{align}
\bar{\bm A} = \begin{bmatrix} \bar{\bm a}(\mu_1) & \cdots & \bar{\bm a}(\mu_d) \end{bmatrix} \in\compl^{M \times d},
\label{arrayphase}
\end{align}
where the steering vectors $\bar{\bm a}(\mu_i),~i=1,\ldots,d,$ are defined as
\begin{align} 
\bar{\bm a}(\mu_i) = \begin{bmatrix} \e^{-\j\frac{(M-1)}{2}\mu_i} & \cdots &\e^{\j\frac{(M-1)}{2}\mu_i} 
\end{bmatrix}. 
\end{align}
After inserting \eqref{arrayphase} into the expression \eqref{steermatphase}, it is once more 
apparent that if the phase reference is at the array centroid, we have $\delta = 0$ and consequently 
$\bm \Delta = \bm I_d$. Moreover, if the phase reference is at the first element, we have $\delta = 
(M-1)/2$. 
\subsection{NC CRB for Two Closely-Spaced Sources}
The result obtained by simplifying the NC CRB for two closely-spaced sources can be summarized in the 
following theorem:
\begin{thm}
For the case of an $M$-element ULA (1-D) and two closely-spaced strictly non-circular sources ($d=2$), 
the deterministic NC Cram\'{e}r-Rao bound can be simplified to expression \eqref{nccrb_two} \red{below}. 
In \eqref{nccrb_two}, we have defined $\Delta\mu = |\mu_2 - \mu_1|$ and $\Delta\phi 
= \Delta\varphi + \delta \Delta\mu$ with $\Delta\varphi = |\varphi_2 - \varphi_1|$. Moreover, 
$\hat{\varrho}_i = N \hat{P}_i / \sigman^2,~i=1,2$ represents the effective SNR of each of 
the two sources.
\begin{figure*}[bp]
\hrule 
\begin{align}
&\traceof{\nc{\bm C}} \approx 50400 \cdot
\bigg(\hat{\rho}^2 \Delta\mu^2 M(M - 1)(M - 2)(M + 2)(M + 1) \Big( \Delta\mu^2  (M - 3)(M + 3) 
\cdot \cos^2(\Delta \phi) + 140 \cdot \sin^2(\Delta \phi) \Big) \bigg. \notag \\
&\quad\quad\quad\quad \bigg. + \left(1 - \hat{\rho}^2\right) M(M - 1)(M + 1) \Big(140 \cdot \Delta\mu^2 (M - 2)(M + 2) 
\cdot \cos^2(\Delta \phi) + 8400 \cdot \sin^2(\Delta \phi) \Big) \bigg)^\inv \cdot 
\frac{\hat{\varrho}_1 + \hat{\varrho}_2}{\hat{\varrho}_1 \hat{\varrho}_2}.
\label{nccrb_two}
\end{align}
\end{figure*}
\label{thm:nccrbtwo}
\end{thm}
%
\begin{IEEEproof}
The proof is given in Appendix \ref{app:nccrbtwo}.
\end{IEEEproof}
It is worth highlighting that the analytical expression in \eqref{nccrb_two} is only an approximate 
result as the derivation involves a Taylor series approximation for small $\Delta \mu$, where the 
higher order terms beyond $\mathcal O(\Delta \mu^4)$ have been neglected. \red{Therefore, \eqref{nccrb_two} 
becomes accurate if $\Delta \mu$ is small.}

Also, note that the behavior of the simplified NC CRB in \eqref{nccrb_two} is symmetric in $\Delta 
\varphi$ as the two sources can be interchanged. Moreover, as any real-valued data stream can be 
multiplied by the factor $-1$, which represents a phase shift of $\pi$, it is also $\pi$-periodic. 
Combining these two results, only the interval $\Delta \varphi \in [0, \pi/2]$ must be considered 
and the general behavior of the NC CRB can be extracted from this interval by mirroring and 
periodification. Consequently, the maximum phase separation is given by $\Delta \varphi = \pi/2$.

Based on the result in \eqref{nccrb_two}, simplified expressions for several special cases can be 
deduced, e.g., for two uncorrelated ($\hat{\rho} = 0$) \red{or} coherent ($\hat{\rho} = 1$) sources as well 
as for $\Delta\phi = 0$ \red{or} $\Delta\phi = \pi/2$. 

\textit{Remark 1:}
One specific case that is worth highlighting is the case $\hat{\rho} = 0$ and $\Delta\phi = \pi/2$, where $\Delta\varphi 
= \pi/2$ and $\delta = 0$. Under these conditions, the NC CRB for two sources in \eqref{nccrb_two} simplifies 
to 
\begin{align}
\traceof{\nc{\bm C}} & \approx \frac{6}{M(M^2 - 1)} \cdot \frac{\hat{\varrho}_1 + 
\hat{\varrho}_2}{\hat{\varrho}_1\hat{\varrho}_2},
\label{nc_decouple}
\end{align}
which is independent of $\Delta\mu$. As \eqref{nc_decouple} resembles the expression for a single source in 
\eqref{nccrb_single}, it is apparent that the individual NC CRB for each of the two sources represents the 
NC CRB for the single source case discussed in the previous section. Hence, the two sources entirely decouple 
as if each of them was present alone. 

\textit{Remark 2:}
Another special case occurs when the two sources approach each other, i.e., $\Delta\mu$ approaches zero. In the
CRB for arbitrary sources this always implies that the CRB tends to infinity. This is, however, not always true 
for the NC CRB. The limit can be computed as
\begin{align}
	\lim_{\Delta \mu \rightarrow 0} \traceof{\nc{\bm C}} & = \frac{1}{1 - \hat{\rho}^2} \cdot 
	\frac{6}{M(M^2 - 1)} \notag \\
	&\quad\quad\quad \cdot \frac{1}{\sin^2(\Delta\phi)} \cdot \frac{\hat{\varrho}_1 + 
	\hat{\varrho}_2}{\hat{\varrho}_1\hat{\varrho}_2}.
	\label{limit}
\end{align}
Thus, for $\hat{\rho} < 1$ and $\Delta\phi > 0$, a finite value is reached. If we have $\hat{\rho} = 0$ and $\Delta\phi = \pi/2$, 
the limit \eqref{limit} corresponds to \eqref{nc_decouple}, and for $\hat{\rho} = 1$ and $\Delta\phi = 0$, the limit 
tends to infinity as the NC CRB matches the CRB.
\subsection{CRB for Two Closely-Spaced Sources}
The corresponding expression of the simplified CRB for two closely-spaced sources is stated as follows:
\begin{thm}
For the case of an $M$-element ULA (1-D) and two closely-spaced sources ($d=2$), the deterministic 
Cram\'{e}r-Rao bound can be simplified to expression \eqref{crb_two} \red{below}.
\begin{figure*}[bp]
\begin{align}
\traceof{\bm C} & \approx 50400 \cdot
\bigg(\hat{\rho}^2 \Delta\mu^2 M(M - 1)(M - 2)(M + 2)(M + 1) \Big( \Delta\mu^2  (M - 3)(M + 3) 
\cdot \cos^2(\Delta \phi) + 140 \cdot \sin^2(\Delta \phi) \Big) \bigg. \notag \\
&\bigg. + 140 \cdot \left(1 - \hat{\rho}^2\right) \Delta\mu^2 M(M - 1)(M - 2)(M + 2)(M + 1) \bigg)^\inv 
\cdot \frac{\hat{\varrho}_1 + \hat{\varrho}_2}{\hat{\varrho}_1 \hat{\varrho}_2}.
\label{crb_two}
\end{align}
\end{figure*}
\label{thm:crbtwo}
\end{thm}
%
\begin{IEEEproof}
The proof is given in Appendix \ref{app:crbtwo}.
\end{IEEEproof}
In analogy to the result for the NC CRB, \eqref{crb_two} \red{becomes exact for small $\Delta \mu$ and} the higher 
order terms beyond $\mathcal O(\Delta\mu^4)$ of the Taylor series expansion are negligible. 

Again, more simplified expressions for several special cases can be derived from \eqref{crb_two}, e.g., 
$\hat{\rho} = 0$, $\hat{\rho} = 1$, $\Delta\phi = 0$, \red{or} $\Delta\phi = \pi/2$. 

\textit{Remark 3:}
A very interesting property of the CRB can be shown for $\hat{\rho} = 1$ and $\Delta\phi = \pi/2$ with 
$\delta = 0$. For these parameters, we can reduce the CRB in \eqref{crb_two} to
\begin{align}
	\traceof{\bm C} & \approx \frac{1}{\Delta\mu^2} \cdot \frac{360}{M(M-1)(M-2)(M+2)(M+1)} \notag \\
	& \quad \cdot \frac{\hat{\varrho}_1 + \hat{\varrho}_2}{\hat{\varrho}_1 \hat{\varrho}_2},
\label{nc_rho}
\end{align}
which corresponds to the expression of the CRB for $\hat{\rho} = 0$ and \red{arbitrary} $\Delta\phi$. This implies 
that a rotation phase separation of $\pi/2$ decorrelates two coherent sources.

\textit{Remark 4:}
In contrast to the NC CRB, the limit for the CRB is given by
\begin{align}
	\lim_{\Delta \mu \rightarrow 0} \traceof{\bm C} = \infty \quad \forall \; \hat{\rho}, \;\; \forall \; \Delta \phi.
\end{align}
Therefore, the NC CRB for strictly non-circular sources exhibits substantial benefits compared to the CRB 
if the sources are closely-spaced, incoherent, and have a non-vanishing phase discrimination $\Delta\phi$.  
\subsection{Analytical NC Gain for Two Closely-Spaced Sources}
Based on the simplified expressions for the two-source case of the NC CRB in \eqref{nccrb_two} and the 
CRB in \eqref{crb_two}, we can explicitly compute the NC gain for two sources as given in \eqref{ncgain} 
\red{below}.
\begin{figure*}[bp]
\hrule 
\begin{align}
\nc{\eta} & = \frac{\traceof{\bm C}}{\traceof{\nc{\bm C}}} \approx
1 + \bigg(140 \cdot (1 - \hat{\rho}^2) M(M - 1)(M + 1) \cdot \sin^2(\Delta\phi)\Big(60 - \Delta\mu^2 (M - 2)(M + 2)\Big)\bigg)\Big/ \notag \\
&\bigg(\Delta\mu^2 M(M - 1)(M - 2)(M + 2)(M + 1) \Big( \Delta\mu^2 \hat{\rho}^2 (M - 3)(M + 3) \cdot \cos^2(\Delta\phi) + 140 \cdot (1 - \hat{\rho}^2 \cos^2(\Delta\phi))\Big)\bigg)
\label{ncgain}
\end{align}
\end{figure*}
As the derivation of \eqref{ncgain} is based on \eqref{nccrb_two} and \eqref{crb_two}, \red{it becomes accurate for 
small source separations $\Delta \mu$ as well.}
We can now analyze the properties of the NC gain \red{expression} for different \red{values} of $\hat{\rho}$, $\Delta\varphi$, and $\delta$.

\textit{Remark 5:}
As already established earlier for an arbitrary number of sources, the NC CRB becomes equal to the CRB if either 
$\hat{\rho} = 1$ or if $\Delta\phi = 0$, where $\Delta\varphi = 0$ and $\delta = 0$. This behavior also reflects in the 
NC gain computed for two strictly non-circular sources as it can easily be verified that for these parameter values, 
the expression \eqref{ncgain} evaluates to $\nc{\eta} = 1$. Hence, no NC gain is obtained in these cases. Note, 
however, that if $\delta \neq 0$, i.e., the phase reference is not at the array centroid, there may be an NC gain 
even if $\Delta\varphi = 0$. 

\textit{Remark 6:}
By analyzing the NC CRB for two closely-spaced sources, we have found that for $\hat{\rho} = 0$ and $\Delta\phi = \pi/2$ 
with $\delta = 0$, the two sources entirely decouple. Evaluating the NC gain expression for these parameters leads 
to
\begin{align}
	\nc{\eta} & \approx \frac{1}{\Delta\mu^2} \cdot \frac{60}{(M-2)(M+2)}. 
\label{nc_gain_decouple}
\end{align}
Thus, this case represents the largest achievable gain for two closely-spaced strictly non-circular sources. It is apparent 
that the NC gain in \eqref{nc_gain_decouple} decays in proportion to $M^{-2}$ but increases as $\Delta\mu$ decreases.

\textit{Remark 7:}
The limit of the NC gain for $\Delta\mu$ approaching zero is given by
\begin{align}
	\lim_{\Delta \mu \rightarrow 0} \nc{\eta} = \infty \quad \forall \; \hat{\rho}, \;\; \forall \; \Delta \phi.
\end{align}
Therefore, the NC gain can theoretically approach infinity if the source separation tends to zero.
\subsection{Two Groups of Equal Phases}
This subsection represents a generalization of the case of two uncorrelated strictly non-circular 
sources \red{to two groups of equal phases}. Let $d$ mutually uncorrelated sources with unit power, i.e., $\hat{\bm R}_{S_0} = \bm I_d$, 
have the phase angles
\begin{align*}
\varphi_i = \varphi^{[1]} + k_i \cdot \pi \quad  \textrm{or} \quad \varphi_i = \varphi^{[2]} + k_i \cdot \pi,~i=1,\ldots,d,
\end{align*}
where $k_i \in \integer$, i.e., modulo $\pi$ there are only two different phase angles: 
$\varphi^{[1]}$ and $\varphi^{[2]}$. Without loss of generality, we can reorder the sources such 
that the $d_1$ sources with phase $\varphi^{[1]}$ are the sources $1, 2, \ldots, d_1$ and the 
remaining $d-d_1$ sources $d_1+1, d_1+2, \ldots, d$ have phase $\varphi^{[2]}$. Thus, the sources 
fall into two groups, where the NC gain depends on the phase separation $|\varphi^{[2]} - \varphi^{[1]}|$ 
of the groups.

Now, in the special case $|\varphi^{[2]} - \varphi^{[1]}| = \pi/2$, i.e., the phase separation 
between the two groups is maximum, it is straightforward to see that the matrices $\ma{G}_0$, 
$\ma{G}_1$, and $\ma{G}_2$ are block diagonal, i.e., they are zero except for the upper left 
$d_1 \times d_1$ block matrix and the lower right $(d-d_1) \times (d-d_1)$ block. Combining 
these matrices and using the fact that the correlation coefficients are zero, we can show 
from the joint CRB that the two groups decouple, that is, the first $d_1$ sources are completely 
decoupled from the remaining $(d-d_1)$ sources. This case can provide a significant gain compared 
to the CRB for arbitrary sources if there are closely-spaced sources that belong
to different groups.
\section{Simulation Results}
\label{sec:simulations}
In this section, we provide simulation results to evaluate the behavior of the $R$-D NC CRB and 
illustrate our analytical results. 
%
\subsection{Behavior of the Deterministic $R$-D NC CRB}
In this subsection, we compare the root mean squared error (RMSE) of the derived deterministic $R$-D NC CRB 
(Det NC CRB) to the deterministic $R$-D CRB (Det CRB) and the existing stochastic $R$-D NC CRB (Sto NC CRB) 
for weak-sense non-circular signals from \cite{delmas2004nccrb}. Moreover, we include the $R$-D NC Standard 
ESPRIT (NC SE) and $R$-D NC Unitary ESPRIT (NC UE) algorithms \cite{steinwandt2014tsp} as well as their non-NC 
counterparts $R$-D Standard ESPRIT (SE) and $R$-D Unitary ESPRIT (UE) \cite{haardt1995unitesprit} into the 
comparison. 
It is assumed that a known number of signals with unit power and real-valued symbols drawn from 
a Gaussian distribution impinge on the array. 

\begin{figure}[t!]
    \centerline{\includegraphics[height=\figheight]{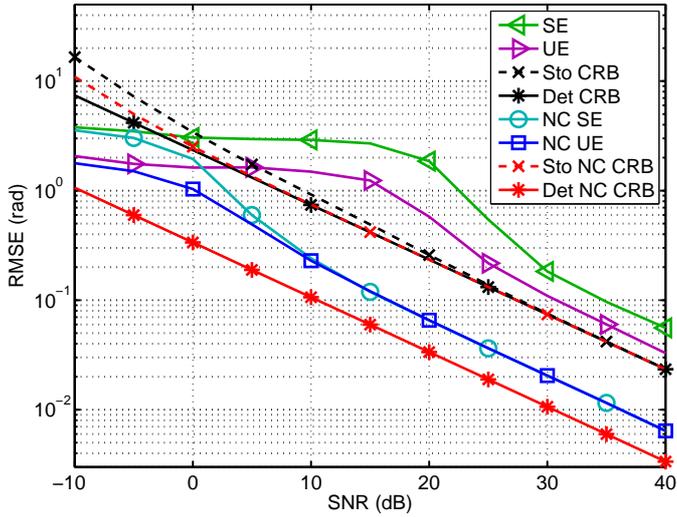}}
    \small
    \caption{Analytical and empirical RMSEs versus SNR for the 12-element 2-D array ($R = 2$) from 
    Fig.~\ref{fig:subsp_2dshinv}, and $N = 20$, $d=3$ correlated sources ($\rho = 0.9$) at $\mu_1^{(1)} = 0.25$, 
    $\mu_2^{(1)} = 0.25$, $\mu_1^{(2)} = 0.5$, $\mu_2^{(2)} = 0.5$, $\mu_1^{(3)} = 0.75$, $\mu_2^{(3)} = 0.75$ 
    with rotation phases $\varphi_1 = 0$, $\varphi_2 = \pi/4$, and $\varphi_3 = \pi/2$.}
    \label{fig:rmse_snr}
\end{figure}

\begin{figure}[t!]
\psfrag{Jxx}{\footnotesize {\color[cmyk]{0,1,1,0} $\bm J_1^{(1)}$}}
\psfrag{J12}{\footnotesize {\color[cmyk]{1,0,1,0} $\bm J_2^{(1)}$}}
\psfrag{J22}{\footnotesize {\color[cmyk]{0,1,1,0} $\bm J_1^{(2)}$}}
\psfrag{J21}{\footnotesize {\color[cmyk]{1,0,1,0} $\bm J_2^{(2)}$}}
\begin{center}
\includegraphics[width=0.7\columnwidth]{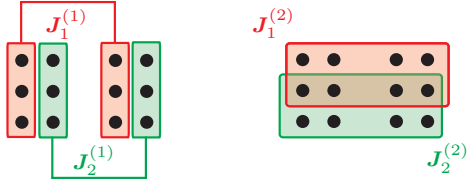}%
\end{center}
\caption{2-D shift invariance for the depicted centro-symmetric $4 \times 3$ sampling grid, left: subarrays
for the first (horizontal) dimension, right: subarrays for the second (vertical)
dimension.}%
\label{fig:subsp_2dshinv}%
\end{figure}

Fig.~\ref{fig:rmse_snr} illustrates the RMSE \red{over all sources} versus the SNR for the centro-symmetric 
2-D array \green{($R=2$)} in Fig.~\ref{fig:subsp_2dshinv} with $M=12$, where $N = 20$ available \red{snapshots} of $d = 3$ 
sources with the spatial frequencies $\mu_1^{(1)} = 0.25$, $\mu_2^{(1)} = 0.25$, $\mu_1^{(2)} = 0.5$, 
$\mu_2^{(2)} = 0.5$, $\mu_1^{(3)} = 0.75$, and $\mu_2^{(3)} = 0.75$, and a real-valued pair-wise correlation 
of $\rho = 0.9$. The rotation phases contained in $\bm \Psi$ are given by $\varphi_1 = 0$, $\varphi_2 = \pi/4$, 
and $\varphi_3 = \pi/2$. It is apparent from Fig.~\ref{fig:rmse_snr} that the NC SE and NC UE algorithms 
perform close to the derived Det NC CRB and that all of these outperform the Sto NC CRB from \cite{delmas2004nccrb}. 

In Table~\ref{table}, we analyze \red{\green{the Det 1-D CRB and the Det 1-D NC CRB} for a varying number of sources $d$ in case of 
a ULA with $M = 4$, $N=20$, and $\mathrm{SNR} = 10$ dB.} The spatial frequencies \green{$\mu_i,~\forall i$} are distributed equally in the 
interval $[-2,2]$ and the rotation phases \green{$\varphi_i,~\forall i$} are drawn randomly. \red{It can be seen that \green{$d_{\rm max}=M-1$} for the CRB and 
\green{$\nc{d_{\rm max}}=2(M-1)$} for the NC CRB are the largest numbers of $d$ that lead to an invertible Fisher matrix, otherwise, 
the problem is ill-posed. Therefore, twice as many sources can be resolved from the strictly non-circular data 
model.}
\subsection{Analytical Results}
In this subsection, \green{we compare the analytical results ``ana'' in \eqref{nccrb_two} and \eqref{crb_two} to the empirical ones 
``emp'' in \eqref{crb} and Corollary \ref{cor:rdnccrb}} obtained by averaging over 1000 Monte-Carlo trials. We have $d=2$ sources that 
impinge on a ULA \green{(1-D)} with the powers $P_1 = 0.5$ and $P_2 = 1.5$. 
The symbols \red{$\bm S_0$ are randomly} drawn from a real-valued Gaussian distribution. 

\begin{figure}[t!]
    \centerline{\includegraphics[height=\figheight]{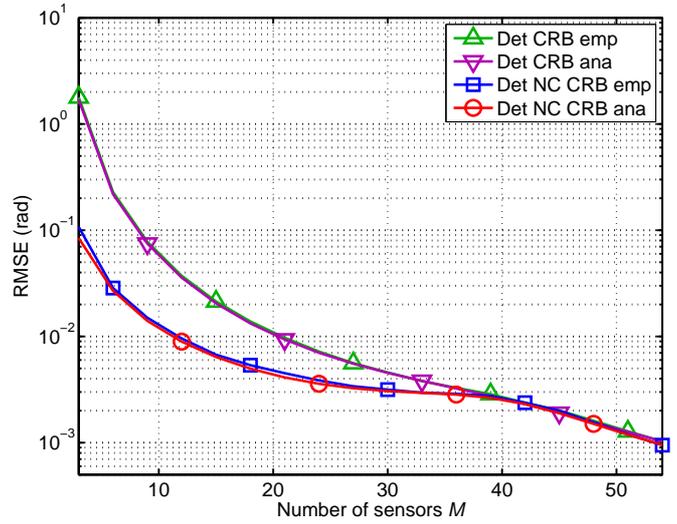}}
    \small
    \caption{Analytical and empirical RMSEs versus the number of sensors $M$ for $d=2$ correlated sources with 
    $N = 10$, $\Delta\mu = 0.1$ rad, $\rho = 0.8$, $\Delta\varphi = \pi/3$, $\delta = (M-1)/2$, $P_1 = 1.5$, 
    $P_2 = 0.5$, and $\sigman^2 = 0.032$.}
    \label{fig:rmse_sensors}
\end{figure}

In Fig.~\ref{fig:rmse_sensors}, we display the RMSE of the Det \green{1-D} NC CRB and \red{the} Det \green{1-D} CRB for $d = 2$ sources 
as a function of the number of sensors $M$, where the square root of the analytical expressions is taken. The 
source separation is $\Delta\mu= 0.1~\mathrm{rad}$ \red{with $\mu_1 = 0$ and $\mu_2 = 0.1$}, however, the actual 
positions are irrelevant and have no impact on the performance. The remaining parameters are given by $N = 10$, 
$\Delta\varphi = \pi/3$, $\delta = (M-1)/2$, \red{i.e., the phase reference is located at the first 
sensor element}, and $\sigman^2 = 0.032$. \green{Moreover, the correlation coefficient $\rho$ is set to $\rho = 0.8$.} 
It is evident that the analytical results agree well with the empirical estimation errors and that both CRBs perform 
similarly for large $M$.

\begin{figure}[t!]
    \centerline{\includegraphics[height=\figheight]{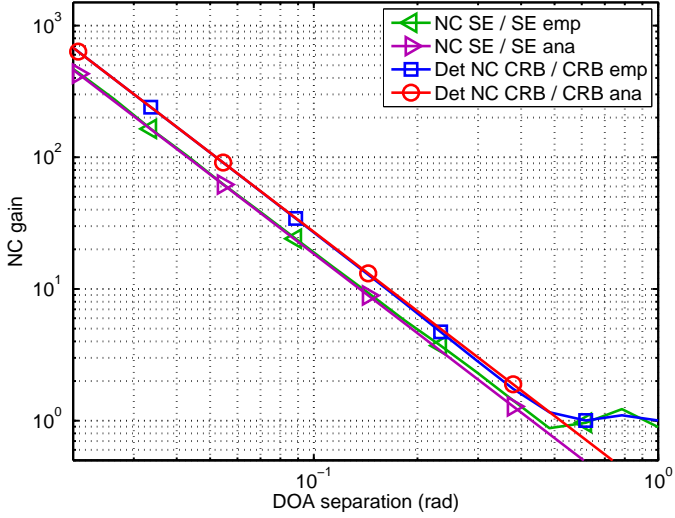}}
    \small
    \caption{Analytical and empirical NC gain versus the source separation $\Delta\mu$ for $d=2$ uncorrelated 
    sources with $M = 15$, $N = 10$, $\Delta\varphi = \pi/2$, $\delta = 0$, $P_1 = 1.5$, $P_2 = 0.5$, and 
    $\sigman^2 = 0.032$.}
    \label{fig:ncgain_sep}
\end{figure}

Fig.~\ref{fig:ncgain_sep} illustrates the asymptotic NC gain in \eqref{ncgain} for $d = 2$ sources as a function 
of $\Delta\mu$. The number of sensors is fixed to $M=15$ and we have $\rho = 0$, $\Delta\varphi = \pi/2$ as well 
as $\delta = 0$. For comparison purposes, we have also included the curves for the analytical NC gain of NC SE 
from \cite{steinwandt2014twosource} for this specific scenario. \green{It can be seen that the NC gain expression becomes 
accurate for small $\Delta\mu$} and that it is largest when $\Delta\mu$ goes to zero. Furthermore, the NC gain of NC SE is close to 
the maximum achievable NC gain computed from the NC CRB.
\begin{table}[!t]
\renewcommand{\arraystretch}{1.3}
\caption{RMSE for a varying number of sources with $M=4$}
\vspace{-0.5em}
\label{table}
\scriptsize
\centering
\begin{tabular}{|c|c|c|c|c|c|c|c|}
\hline
RMSE & $d=1$ & $d=2$ & $d=3$ & $d=4$ & $d=5$ & $d=6$ & $d=7$\\
\hline\hline 
CRB & $0.02$ & $0.13$ & $0.80$ & $\infty$ & $\infty$ & $\infty$ & $\infty$ \\
\hline 
NC CRB & $0.02$ & $0.11$ & $0.12$ & $0.14$ & $0.35$ & $2.93$ & $\infty$\\ 
\hline
\end{tabular}
\end{table}
\section{Conclusion}
\label{sec:conclusions}
In this paper, we have presented a closed-form expression of the deterministic $R$-D NC Cram\'{e}r-Rao 
bound for multi-dimensional strictly non-circular (rectilinear) signals. This bound serves as a benchmark 
for the recently developed algorithms, e.g., $R$-D NC Standard ESPRIT and $R$-D NC Unitary ESPRIT, that 
exploit the NC structure of such strictly non-circular signals and thus outperform the traditional methods 
for arbitrary signals. Based on the resulting $R$-D NC CRB expression and assuming the $R$-D array to be 
separable and centro-symmetric, we have shown that in the special cases of equal phases and full coherence 
of the strictly non-circular signals as well as for a single strictly non-circular source, the deterministic 
$R$-D NC CRB reduces to the existing deterministic $R$-D CRB for arbitrary signals. This suggests that no 
NC gain can be achieved in these specific cases.
Furthermore, we have simplified the derived NC CRB and the existing CRB for the special case of two 
closely-spaced strictly non-circular signals captured by a uniform linear array (ULA). With these 
simplified CRB expressions, we have then analytically computed the maximum achievable asymptotic NC 
gain for this scenario. The resulting expression only depends on the various physical parameters, e.g., 
the number of sensors, the signal correlation, etc. Additionally, we have analyzed the dependence of 
the NC gain on these parameters to find \red{that the largest NC gain is obtained if the two sources are 
closely-spaced, incoherent, and have a non-vanishing phase discrimination. }
\appendices
%
\section{Proof of Theorem \ref{thm:nccrb}} 
\label{app:nccrb}
For convenience, we start our derivation by vectorizing the $R$-D NC data model in \eqref{model2} \red{by 
using the property $\mathrm{vec}\{\bm A \bm X \bm B\} = (\bm B^\trans \otimes \bm A) \cdot \mathrm{vec}\{\bm X\}$ 
for arbitrary matrices $\bm A$, $\bm B$, and $\bm X$ of appropriate sizes. We obtain}
\begin{equation}
\bm x = \vecof{\bm X} = \left(\bm I_N \otimes \bm A \bm \Psi\right) \bm s_0 + \bm n \in\compl^{MN \times 1}, 
\label{model4}
\end{equation}
where $\bm s_0 = \vecof{\bm S_0} = [\bm s_0^\trans(1), \ldots, \bm s_0^\trans(N)]^\trans \in\real^{Nd 
\times 1}$ with $\bm s_0(t),~t=1,\ldots,N$, being the $t$-th column of $\bm S_0$, and $\bm n = \vecof{\bm N} 
\in\compl^{MN \times 1}$. To suit the deterministic data assumption, the signal vector $\bm s_0$ is 
assumed to be deterministic and unknown to the receiver, while the sensor noise $\bm n$ is zero-mean 
circularly symmetric white complex Gaussian distributed, i.e., $\mathbb E\{\bm n \bm n^\trans\} = 
\bm 0$. Hence, the observations $\bm x$ satisfy the model  
\begin{equation}
\bm x \sim \mathcal{CN}(\bm \nu,\bm \Sigma), 
\label{statmodel}
\end{equation}
where $\bm \nu = \left(\bm I_N \otimes \bm A \bm \Psi\right) \bm s_0$ and $\bm \Sigma = \sigman^2\bm I_{MN}^{}$ 
are the mean and the covariance of the array output vector $\bm x$.

Let us now define the real-valued vector of unknown parameters as
\begin{equation}
\bm \xi = \begin{bmatrix} \bm \mu^\trans & \bm s_0^\trans & \bm \varphi^\trans & \sigman^2 
\end{bmatrix}^\trans \in\real^{[(R + N + 1)d + 1] \times 1}. 
\label{paravec}
\end{equation}
Here, $\bm \mu = [\RdoneT{\bm \mu}, \ldots, \RdT{\bm \mu}]^\trans \in\real^{Rd \times 1}$ is 
the principal parameter vector of interest and $\bm s_0 \in\real^{Nd \times 1}$, $\bm \varphi \in 
\real^{d \times 1}$, and $\sigman^2$ are the nuisance parameters. As the CRB matrix is usually computed 
by taking the inverse of the Fisher information matrix (FIM) $\bm J$, we first need to calculate $\bm J$. 
Due to \eqref{statmodel}, i.e., $\bm x$ is Gaussian distributed, the Slepian-Bangs formulation 
\cite{stoica2005spectral} of the FIM is still valid for the strictly non-circular data model in 
\eqref{model4}. Hence, the Slepian-Bangs formulation of $\bm J$ for the parameter vector $\bm \xi$ is 
given by \cite{stoica2005spectral}
\begin{eqnarray}
\begin{aligned}
\bm J_{p,q} &= \mathrm{Tr}\left\{\bm \Sigma^\inv\frac{\partial \bm \Sigma}{\partial \bm \xi_p}
\bm \Sigma^\inv \frac{\partial \bm \Sigma}{\partial \bm \xi_q}\right\} \\
&\quad\quad\quad\quad\quad + 2 \cdot \mathrm{Re}\left\{\left(\frac{\partial \bm \nu}{\partial \bm \xi_p}\right)^\herm
\bm \Sigma^\inv \frac{\partial \bm \nu}{\partial \bm \xi_q}\right\},
\label{fim}
\end{aligned}
\end{eqnarray}
\vspace{-0.5em}
\begin{equation}
\qquad\qquad\qquad~p,q=1,\ldots,(R + N + 1)d+1. \notag
\end{equation}
Note that we are only interested in the CRB for $\bm \mu$, denoted as $\nc{\bm C}$. Therefore, it 
is sufficient to compute the upper left block of $\bm J^\inv$. In order to find $\bm J$ from \eqref{fim}, 
the partial derivatives of $\bm \nu$ with respect to the parameters of $\bm \xi$ can be calculated 
straightforwardly. We have 
\begin{align} 
\frac{\partial \bm \nu}{\partial \bm \mu^\trans} 
& = \left(\bm I_N \otimes (\bm D (\bm I_R \otimes \bm \Psi)) \right) \Rd{\tilde{\bm S}}_0 \in\compl^{MN \times Rd}, 
\label{mu_theta}
\end{align}
where $\bm D$ is given in \eqref{Drd} and $\Rd{\tilde{\bm S}}_0 = [(\bm I_R \otimes \tilde{\bm S_0}(1)), \ldots, 
(\bm I_R \otimes \tilde{\bm S_0}(N))]^\trans \in\real^{NRd \times Rd}$ with $\tilde{\bm S_0}(t) = \diagof{\bm s_0(t)} 
\in\real^{d \times d}$. For the remaining parameters, we get 
\begin{align}
\frac{\partial \bm \nu}{\partial \bm s_0^\trans} & = \left(\bm I_N \otimes \bm A \bm \Psi \right) 
\in\compl^{MN \times Nd}, \notag \\
\frac{\partial \bm \nu}{\partial \bm \varphi^\trans} & = \j \left(\bm I_N \otimes \bm A \bm \Psi \right) 
\tilde{\bm S}_0 \in\compl^{MN \times d}, ~~
\frac{\partial \bm \nu}{\partial \sigman^2} = \bm 0 \in\real^{MN \times 1}, \notag   
\end{align}
where $\tilde{\bm S}_0 = [\tilde{\bm S_0}(1), \ldots, \tilde{\bm S_0}(N)]^\trans \in\real^{Nd \times d}$. 
Next, these results are combined to obtain 
\begin{align}
\frac{d \bm \nu}{d \bm \xi^\trans} 
& = \left[ \begin{matrix} \left(\bm I_N \otimes (\bm D (\bm I_R \otimes \bm \Psi)) \right) \Rd{\tilde{\bm S}}_0 & 
\left(\bm I_N \otimes \bm A \bm \Psi \right), \end{matrix} \right. \notag\\
&\quad~\left. \begin{matrix} \j \left(\bm I_N \otimes \bm A \bm \Psi \right) 
\tilde{\bm S}_0 & \bm 0 \end{matrix} \right] \in\compl^{MN \times [(R + N + 1)d+1]}, 
\label{mud}
\end{align}
As for the derivative of $\bm \Sigma$ with respect to $\bm \xi$, the only non-zero term is 
\begin{align}
\frac{d \bm \Sigma}{d \sigman^2} = \bm I_{MN},  
\label{gammadsigma}
\end{align}
such that 
\begin{align}
\frac{d \bm \Sigma}{d \bm \xi^\trans} = \begin{bmatrix} \bm 0 & \bm 0 & \bm 0 & \bm I_{MN} \end{bmatrix} 
\in\compl^{MN \times [(R + N + 1)d + MN]}. 
\label{gammad}
\end{align}
Inserting \eqref{mud} and \eqref{gammad} into \eqref{fim} and bearing in mind that we are 
interested in the $\bm \mu$-block of $\bm J$, only the second term of \eqref{fim} is of concern. 
Therefore, we only consider the non-zero block $\tilde{\bm J}$ of $\bm J$, which is given by 
\begin{align}
\tilde{\bm J} = \begin{bmatrix} \bm J_{\bm \mu,\bm \mu} & \bm J_{\bm \mu,\bm s_0} & \bm J_{\bm \mu,\bm \varphi} \\
\bm J_{\bm s_0,\bm \mu} & \bm J_{\bm s_0,\bm s_0} & \bm J_{\bm s_0,\bm \varphi} \\
\bm J_{\bm \varphi,\bm \mu} & \bm J_{\bm \varphi,\bm s_0} & \bm J_{\bm \varphi,\bm \varphi} \end{bmatrix}
= \frac{2}{\sigman^2} \cdot \mathrm{Re}\left\{\bm G^\herm\bm G\right\}, 
\label{J}
\end{align}
where 
\begin{align}
\bm G & = \left[ \begin{matrix} \left(\bm I_N \otimes (\bm D (\bm I_R \otimes \bm \Psi)) \right) \Rd{\tilde{\bm S}}_0 
& \left(\bm I_N \otimes \bm A \bm \Psi \right), \end{matrix} \right. \notag \\
&\quad\quad\quad\quad\quad \left. \begin{matrix} \j \left(\bm I_N \otimes \bm A \bm \Psi \right) 
\tilde{\bm S}_0 \end{matrix} \right] \in\compl^{MN \times (R + N + 1)d}.
\label{Gmat}
\end{align}
It is easy to see that $\tilde{\bm J} = \tilde{\bm J}^\trans$. Consequently, only the block matrices on and above the diagonal of 
$\tilde{\bm J}$ need to be computed. For the block matrix $\bm J_{\bm \mu, \bm \mu}$, we obtain
\begin{align}
\bm J_{\bm \mu, \bm \mu} & = \frac{2}{\sigman^2} \cdot \sum_{t=1}^N \realof{ (\bm I_R \otimes \tilde{\bm S}_0(t)) 
(\bm I_R \otimes \bm \Psi^\conj) \bm D^\herm \right. \notag\\
& \left. \quad\quad\quad\quad\quad\quad\quad\quad \cdot \bm D (\bm I_R \otimes \bm \Psi) (\bm I_R \otimes \tilde{\bm S}_0(t))} \\
& = \frac{2}{\sigman^2} \cdot \mathrm{Re}\Bigg\{ \left((\bm I_R \otimes \bm \Psi^\conj) \bm D^\herm \bm D 
(\bm I_R \otimes \bm \Psi)\right) \notag \Bigg. \\ 
&\left. \quad\quad\quad\quad\quad \odot \sum_{t=1}^N (\bm 1_R \otimes \bm s_0(t))
(\bm 1_R \otimes \bm s_0(t))^\trans \right\} \\
& = \frac{2}{\sigman^2} \cdot \mathrm{Re}\Bigg\{ \left((\bm I_R \otimes \bm \Psi^\conj) \bm D^\herm \bm D 
(\bm I_R \otimes \bm \Psi)\right) \notag \Bigg. \\ 
&\left. \quad\quad\quad\quad\quad\quad\quad \odot \left( \bm 1_{R \times R} \otimes \sum_{t=1}^N \bm s_0(t) \bm s^\trans_0(t) \right) \right\} \\
& = \frac{2N}{\sigman^2} \cdot \mathrm{Re}\Bigg\{ \left((\bm I_R \otimes \bm \Psi^\conj) \bm D^\herm \bm D 
(\bm I_R \otimes \bm \Psi)\right) \notag \Bigg. \\ 
&\left. \quad\quad\quad\quad\quad\quad\quad \odot \left( \bm 1_{R \times R} \otimes \frac{1}{N} \bm S^{}_0 \bm S_0^\trans \right) \right\} \\
& = \frac{2N}{\sigman^2} \cdot \bm G_2 \odot \Rd{\hat{\bm R}}_{S_0} \in\real^{Rd \times Rd}, 
\label{Jtheta}
\end{align}
where $\bm G_2$ is defined according to \eqref{g2rd} and we have used the fact that $\diagof{\bm a} \bm C 
\diagof{\bm b} = \bm C \odot (\bm a \bm b^\trans)$ for arbitrary vectors $\bm a \in\compl^{M}, \bm b 
\in\compl^{N}$, and a matrix $\bm C \in\compl^{M \times N}$. In a similar manner, the other blocks of 
$\tilde{\bm J}$ can be computed. The results are given by
\begin{align}
\bm J_{\bm s_0, \bm s_0} & = \frac{2}{\sigman^2} \cdot \bm I_N \otimes \bm G_0 \in\real^{Nd \times Nd} \\
\bm J_{\bm \varphi, \bm{\varphi}} & = \frac{2N}{\sigman^2} \cdot \bm G_0 \odot \hat{\bm R}_{S_0} \in\real^{d \times d} \\
\bm J_{\bm \mu, \bm s_0} & = \frac{2}{\sigman^2} \cdot \RdT{\tilde{\bm S}}_0  
\left(\bm I_N \otimes \bm G_1\right) \in\real^{Rd \times Nd} \\
\bm J_{\bm s_0, \bm \varphi} & = - \frac{2}{\sigman^2} \cdot \left(\bm I_N \otimes \bm H_0 \right) \tilde{\bm{S}_0} \in\real^{Nd \times d}\\ 
\bm J_{\bm \mu, \bm \varphi} & = - \frac{2N}{\sigman^2} \cdot \bm H_1 \odot (\bm 1_R \otimes \bm{\hat{R}}_{S_0}) \in\real^{Rd \times d},
\label{Jblocks}
\end{align}
where the matrices $\bm G_n$ and $\bm H_n,~n=0,1,2$ are given in \eqref{g0rd}-\eqref{h1rd}. Note that 
we have the symmetries $\bm G_0 = \bm G_0^\trans$, $\bm G_2 = \bm G_2^\trans$, and $\bm H_0 = -\bm H_0^\trans$. 

In the next step, we need to extract the upper left block of $\tilde{\bm J}^\inv$. To this end, we make use of the 
following lemma:
\begin{lem} 
For matrices 
   $\ma{A} \in \compl^{p \times p}$, $\ma{B} \in \compl^{p \times q}$, $\ma{C} \in \compl^{p \times r}$,
   $\ma{D} \in \compl^{q \times p}$, $\ma{E} \in \compl^{q \times q}$, $\ma{F} \in \compl^{q \times r}$,
   $\ma{G} \in \compl^{r \times p}$, $\ma{H} \in \compl^{r \times q}$, and $\ma{J} \in \compl^{r \times r}$
   the upper left $p \times p$ block of the matrix
   \begin{align}
	    \ma{K} = 	\begin{bmatrix}
	                  \ma{A} & \ma{B} & \ma{C} \\
	                  \ma{D} & \ma{E} & \ma{F} \\
	                  \ma{G} & \ma{H} & \ma{J} 
	              \end{bmatrix}^\inv
	 \label{inversion}
   \end{align}
   is given by
   \begin{align}
	    \ma{K}_{1:p, 1:p} & = \Big( \ma{A}  - \ma{B} \ma{E}^\inv \ma{D}  - \ma{B} \ma{E}^\inv \ma{F} \ma{S}_E^\inv \ma{H} \ma{E}^\inv \ma{D} \notag \\
	    & + \ma{B} \ma{E}^\inv \ma{F} \ma{S}_E^\inv \cdot \ma{G} + \ma{C} \ma{J}^\inv \ma{H} \ma{E}^\inv \ma{D} \nonumber \\
	    & + \ma{C} \ma{J}^\inv \ma{H} \ma{E}^\inv \cdot  \ma{F} \ma{S}_E^\inv \ma{H} \ma{E}^\inv \ma{D} 
	     - \ma{C} \ma{S}_E^\inv \ma{G} \Big)^\inv, \notag 
   \end{align}
where $\ma{S}_E  =  \ma{J} - \ma{H} \ma{E}^\inv \ma{F}$.
   \label{lem_blinv3}
   \end{lem}
\begin{IEEEproof}
The proof of Lemma \ref{lem_blinv3} can easily be constructed by applying the inversion formula for a $2 \times 2$ 
block-partitioned matrix \cite{lütke1996matrices} to the $3\times 3$ block matrix in \eqref{inversion} twice.
\end{IEEEproof}

Applying Lemma \ref{lem_blinv3} to compute the upper left block of $\tilde{\bm J}^\inv$, it is straightforward to obtain 
the expression in \eqref{eqn_thekillercrb}, where we have
\begin{align*}
   \ma{S}_E & = \frac{2N}{\sigman^2} \cdot \left( \ma{G}_0 - \ma{H}_0^\trans \ma{G}_0^\inv \ma{H}_0 \right) \odot \Rd{\hat{\bm R}}_{S_0} \\
   \ma{B} \ma{E}^\inv \ma{D} & = \frac{2N}{\sigman^2} \cdot \left( \ma{G}_1 \ma{G}_0^\inv \ma{G}_1^\trans \right) \odot \Rd{\hat{\bm R}}_{S_0} \\
   \ma{B} \ma{E}^\inv \ma{F} & = -\frac{2N}{\sigman^2} \cdot \left( \ma{G}_1 \ma{G}_0^\inv \ma{H}_0 \right) \odot \Rd{\hat{\bm R}}_{S_0} \\
   \ma{H} \ma{E}^\inv \ma{D} & = -\frac{2N}{\sigman^2} \cdot \left( \ma{H}_0^\trans \ma{G}_0^\inv \ma{G}_1^\trans \right) \odot \Rd{\hat{\bm R}}_{S_0} \\
   \ma{H} \ma{E}^\inv \ma{F} & = \frac{2N}{\sigman^2} \cdot \left( \ma{H}_0^\trans \ma{G}_0^\inv \ma{H}_0 \right) \odot \Rd{\hat{\bm R}}_{S_0}.
\end{align*}
This concludes the proof. \qed
\section{}
\label{app:real}
In this section, we prove that for $\rd{\delta} = 0~\forall \;r$ and subsequently $\bm A = \bar{\bm A}$, 
the matrices $\bm A^\herm \bm A \in\real^{d \times d}$, $\bm D^\herm \bm A \in\real^{Rd \times d}$, and 
$\bm D^\herm \bm D \in\real^{Rd \times Rd}$ are real-valued.
To this end, we make use of the following lemma:
\begin{lem}
For two arbitrary non-singular left $\bm \Pi$-real matrices $\bm X \in\compl^{M \times N}$ and $\bm Y 
\in\compl^{M \times N}$ satisfying $\bm \Pi \bm X^\conj = \bm X$ and $\bm \Pi \bm Y^\conj = \bm Y$, 
respectively, the following identity holds:
\begin{align}
\bm Y^\herm \bm X & = (\bm \Pi \bm Y^\conj)^\herm \bm \Pi \bm X^\conj = \bm Y^\trans \bm \Pi \bm \Pi \bm X^\conj \notag\\
& = (\bm Y^\herm \bm X)^\conj \in\real^{N \times N}.
\end{align}
\end{lem}
Therefore, to prove that the aforementioned matrices are real-valued, we simply show that the matrices 
$\bm A$ and $\bm D$ are left $\bm \Pi$-real. It is straightforward to see that due to $\rd{\delta} = 0~\forall \;r$, 
this is the case for $\bm A$ (cf. Equation \eqref{centro}). As for the matrix $\bm D$, we utilize the linearity 
of the differentiation operator and obtain
\begin{align}
\bm \Pi \bm D^\conj & = \bm \Pi \left(\frac{\partial \bm A}{\partial \bm \mu} \right)^\conj 
= \frac{\partial \bm \Pi \bm A^\conj}{\partial \bm \mu} = \frac{\partial \bm A}{\partial \bm \mu} = \bm D,
\end{align}
which also renders $\bm D$ left $\bm \Pi$-real and concludes the proof. \qed
\section{Proof of Theorem \ref{thm:nccrbsingle}}
\label{app:nccrbsingle}
Evaluating the $R$-D NC CRB in \eqref{eqn_thekillercrb} for the special case $d=1$, the array steering 
matrix $\bm A$ reduces to $\bm a(\bm \mu)$, $\bm D = [\bm d^{(1)},\ldots,\bm d^{(R)}] 
\in\compl^{M\times R}$, $\bm \Psi = \expof{\j\varphi}$, and $\hat{\bm R}_{S_0} = \bm s_0^\trans 
\bm s_0^{} /N = \hat{P}$, where $\bm s_0 \in\real^{N\times 1}$. \red{Moreover, we choose $\rd{\delta} = 
0~\forall \;r$ for simplicity.} Then, dropping the dependence of $\bm a$ on $\bm \mu$ and utilizing the definitions 
in \eqref{steer} and \eqref{derivrd}, respectively, we have
\begin{align}
\bm a^\herm \bm a^{} & = \prod_{r=1}^R \rdH{\bm a} \rd{\bm a} = 
\prod_{r=1}^R M_r = M, \label{real1}\\
\rdH{\bm d} \bm a & = \prod_{\substack{p = 1 \\ p \neq r }}^R \pdH{\bm a} \pd{\bm a} \cdot 
\rdH{\tilde{\bm d}} \rd{\bm a} \notag\\ 
& = \prod_{\substack{p = 1 \\ p \neq r }}^R \pdH{\bm a} \pd{\bm a} \cdot \left(-\j \sum_{m_r=1}^{M_r} k_{m_r} \right) = 0 ~~\forall \;r, \label{real2}\\
\rdH{\bm d} \rd{\bm d} & = \prod_{\substack{p = 1 \\ p \neq r }}^R \pdH{\bm a} \pd{\bm a} \cdot 
\rdH{\tilde{\bm d}} \rd{\tilde{\bm d}} \notag\\
& = \prod_{\substack{p = 1 \\ p \neq r }}^R \pdH{\bm a} \pd{\bm a} \cdot \left(\sum_{m_r=1}^{M_r} k_{m_r}^2 \right) \notag\\
& = \frac{M}{M_r} \sum_{m_r=1}^{M_r} k_{m_r}^2 = \rd{\Gamma} ~~\forall \;r .
\label{real3}
\end{align}
Using the results in \eqref{real1}-\eqref{real3}, the matrices $\bm G_n$ and $\bm H_n,~n=0,1,2,$ simplify to
\begin{align}
\bm G_0 & = M, \quad\quad \bm G_1 = \bm H_0 = \bm H_1 = 0, \\
\bm G_2 & = \bm D^\herm \bm D = \diagof{\left[ \Gamma^{(1)},\ldots,\Rd{\Gamma} \right]},
\end{align}
where in $\bm G_2$, the terms $\bm d^{(r_1)^\herm} \bm d^{(r_2)}$ for $r_1 \neq r_2$ evaluate to zero due to 
\eqref{real2}.
Inserting these expressions into \eqref{eqn_thekillercrb}, the remaining part of the $R$-D NC CRB matrix 
is given by 
\begin{align}
\nc{\bm C} & = \frac{\sigman^2}{2N \hat{P}} \cdot \left\{ \diagof{\left[ \Gamma^{(1)},\ldots,\Rd{\Gamma} \right]} \right\}^\inv \\
& = \diagof{\left[ {\nc{C}}^{(1)},\ldots, {\nc{C}}^{(R)} \right]} \in\real^{R \times R},
\end{align}
where 
\begin{align}
\rd{{\nc{C}}} = \frac{\sigman^2}{2N \hat{P}} \cdot \frac{M_r}{M} \cdot \frac{1}{\sum_{m_r=1}^{M_r} k_{m_r}^2}~ \forall \; r,
\end{align}
which is the desired result. \qed
\section{Proof of Theorem \ref{thm:nccrbtwo}}
\label{app:nccrbtwo}
Based on the model in \eqref{model3} after inserting \eqref{arrayphase}, we start the proof by assuming 
without loss of generality that the phase reference is at the array centroid, i.e., $\bm \Delta = \bm I_d$ 
such that $\bm A = \bar{\bm A}$ and $\bm \Phi = \bm \Psi$. Using the results from Appendix \ref{app:real}, 
we can write the real-valued matrices $\bm A^\herm \bm A$, $\bm D^\herm \bm A$, and $\bm D^\herm \bm D$ 
as 
\begin{align}
\bm A^\herm \bm A = \begin{bmatrix} M & \alpha \\ \alpha & M \end{bmatrix} \! ,~
\bm D^\herm \bm A = \begin{bmatrix} 0 & \beta \\ -\beta & 0 \end{bmatrix},~
\bm D^\herm \bm D = \begin{bmatrix} \Gamma & \gamma \\ \gamma & \Gamma \end{bmatrix},
\notag
\end{align}
where we have defined $\alpha = \bm a_1^\herm \bm a_2 = \bm a_2^\herm \bm a_1 $, $\beta = \bm d_1^\herm 
\bm a_2 = - \bm d_2^\herm \bm a_1$, and $\gamma = \bm d_1^\herm \bm d_2 = \bm d_2^\herm \bm d_1$. Then, 
the matrices $\bm G_0$ and $\bm H_0$ can be written as 
\begin{align}
\bm G_0 &= \realof{\bm \Psi^\conj \bm A^\herm \bm A \bm \Psi} 
= \begin{bmatrix} M & \alpha \cdot \cos(\Delta\varphi) \\ \alpha \cdot \cos(\Delta\varphi) & M \end{bmatrix} \notag \\
\bm H_0 &= \imagof{\bm \Psi^\conj \bm A^\herm \bm A \bm \Psi} 
= \begin{bmatrix} 0 & \alpha \cdot \sin(\Delta\varphi) \\ -\alpha \cdot \sin(\Delta\varphi) & 0 \end{bmatrix}.
\notag
\end{align}
The matrices $\bm G_1$, $\bm H_1$, and $\bm G_2$ can be expressed in a similar manner. In order to obtain 
an expression of the 1-D NC CRB that only depends on the physical parameters, e.g, $M$, $\rho$, $\Delta 
\varphi$, etc., we approximate the scalars $\alpha$, $\beta$, and $\gamma$ using a Taylor series expansion 
for small source separations $\Delta\mu = |\mu_2 - \mu_1|$. \red{Hence, these approximations become accurate 
for a small $\Delta\mu$.} Therefore, for $\alpha$, we have
\begin{align}
\alpha &= \sum_{m=-\frac{(M-1)}{2}}^{\frac{(M-1)}{2}} \e^{\j m \Delta\mu}
\approx M + \j \Delta\mu \cdot \sum_{m=-\frac{(M-1)}{2}}^{\frac{(M-1)}{2}} m \notag\\
&- \frac{\Delta\mu^2}{2} \cdot \sum_{m=-\frac{(M-1)}{2}}^{\frac{(M-1)}{2}} m^2 - ~ \cdots \notag\\
&\approx M - \frac{M}{24} \Delta\mu^2 (M^2-1) + \mathcal O(\Delta\mu^4).
\notag
\end{align}
Note that the terms containing odd powers of $m$ evaluate to zero. 
\red{Similarly, in case of a small $\Delta\mu$, we get for $\beta$ and $\gamma$ the expressions}
\begin{align}
\beta &= -\j \cdot \sum_{m=-\frac{(M-1)}{2}}^{\frac{(M-1)}{2}} m \cdot \e^{\j m \Delta\mu} \notag \\
&\approx -\j \cdot \sum_{m=-\frac{(M-1)}{2}}^{\frac{(M-1)}{2}} m \cdot \left(1 + \j m \Delta\mu - \frac{\Delta\mu^2}{2} m^2 - ~\cdots \right) \notag \\
&\approx \frac{M}{12} \Delta\mu (M^2-1) - \frac{M}{1440} \Delta\mu^3 (3M^4 - 10M^2+7) + \mathcal O(\Delta\mu^5), \notag \\
\gamma &= \sum_{m=-\frac{(M-1)}{2}}^{\frac{(M-1)}{2}} m^2 \cdot \e^{\j m \Delta\mu} \notag \\
&\approx -\j \cdot \sum_{m=-\frac{(M-1)}{2}}^{\frac{(M-1)}{2}} m^2 \cdot( 1 + \j m \Delta\mu - \frac{\Delta\mu^2}{2} m^2 - ~\cdots) \notag \\
&\approx \frac{M}{12} (M^2-1) - \frac{M}{480} \Delta\mu^2 (3M^4 - 10M^2+7) + \mathcal O(\Delta\mu^4).
\notag
\end{align}
Finally, with the sample covariance matrix
\begin{align}
\hat{\bm R}_{S_0} = \begin{bmatrix} \hat{P}_1 & \hat{\rho} \sqrt{\hat{P}_1 \hat{P}_2} \\ 
\hat{\rho} \sqrt{\hat{P}_1 \hat{P}_2} & \hat{P}_2 \end{bmatrix}
\end{align}
and the help of the Taylor approximation terms above, we can evaluate the 1-D NC CRB expression in Corollary 
\ref{cor:rdnccrb} for two closely-spaced strictly non-circular sources. Due to the cancellation of relevant 
terms when using only approximation terms of lower order, we also need to consider higher-order Taylor 
approximation terms\footnote{\red{Here, we used Taylor approximation terms \green{up to} 
the 6th order.}} for $\alpha$, $\beta$, and $\gamma$. After some tedious calculations, we obtain
\begin{align}
\traceof{\nc{\bm C}} = \frac{\sigman^2}{2N} \cdot z \cdot \frac{\hat{P}_1 + \hat{P}_2}{\hat{P}_1 \hat{P}_2},
\label{nccrb_app}
\end{align}
where
\begin{align}
z = \frac{x_0 + x_1 \Delta\mu^2 + x_2 \Delta\mu^4 + \cdots}
{y_1 \Delta\mu^2 + y_2 \Delta\mu^4 + y_3 \Delta\mu^6 + \cdots}.
\end{align}
It is apparent that the first term in the numerator and the first two terms in the denominator of 
\eqref{nccrb_app} are dominant. Neglecting the non-relevant higher-order terms in the numerator and 
denominator of \eqref{nccrb_app} and applying some algebraic manipulations, an expression in the form 
of \eqref{nccrb_two} can be deduced. Finally, to make the result more general, we consider an arbitrary 
phase reference and substitute $\Delta\varphi$ by $\Delta\phi$ to obtain \eqref{nccrb_two}. This concludes 
the proof. \qed
%
%
\section{Proof of Theorem \ref{thm:crbtwo}}
\label{app:crbtwo}
The proof of Theorem \ref{thm:crbtwo} follows the same steps as the proof in Appendix \ref{app:nccrbtwo}.
Under the same assumptions, we compute the matrices $\bm A^\herm \bm A$, $\bm D^\herm \bm A$, and $\bm D^\herm 
\bm D$ in the same way. The difference is, however, that we evaluate the 1-D CRB expression given in 
\eqref{rdcrb}. Using the same Taylor series approximations as before, we obtain a similar expression as 
\eqref{nccrb_app}. Finally, neglecting the non-dominant terms in the numerator and the denominator, and 
substituting $\Delta\phi$ for $\Delta\varphi$, we arrive at the expression in \eqref{crb_two} to prove 
this theorem. \qed 
%
%
%
\bibliographystyle{IEEEbib}
\bibliography{refs_nc_crb}

\end{document}